\documentclass{aa}
\usepackage{natbib,twoopt}
\usepackage{amsmath}

\usepackage{color}

\usepackage[breaklinks=true]{hyperref}
\makeatletter
  \newcommandtwoopt{\intheteads}[3][][]{\href{http://adsabs.harvard.edu/abs/#3}%
    {\def\hyper@linkstart##1##2{}%
     \let\hyper@linkend\@empty\citealp[#1][#2]{#3}}}
  \newcommandtwoopt{\citepads}[3][][]{\href{http://adsabs.harvard.edu/abs/#3}%
    {\def\hyper@linkstart##1##2{}%
     \let\hyper@linkend\@empty\citep[#1][#2]{#3}}}
  \newcommandtwoopt{\citetads}[3][][]{\href{http://adsabs.harvard.edu/abs/#3}%
    {\def\hyper@linkstart##1##2{}%
     \let\hyper@linkend\@empty\citet[#1][#2]{#3}}}
  \newcommandtwoopt{\citeyearads}[3][][]%
    {\href{http://adsabs.harvard.edu/abs/#3}
    {\def\hyper@linkstart##1##2{}%
     \let\hyper@linkend\@empty\citeyear[#1][#2]{#3}}}
\makeatother

\defcitealias{2025A&A...702A..12R}{RBB25}

\usepackage{graphicx}
\usepackage{txfonts}

\begin{document}

   \title{FRANZ: Framework for analytical one-zone blastwave dynamics}

   \author{Leonard E. C. Romano
          \inst{1}\fnmsep\inst{2}\fnmsep\inst{3}\fnmsep\thanks{Corresponding author: Leonard E. C. Romano\\\email{lromano@usm.lmu.de}}
          }

   \institute{Excellence Cluster ORIGINS, Boltzmannstr. 2, 85748 Garching, Germany
   \and
    European Southern Observatory (ESO), Karl-Schwarzschild-Stra{\ss}e 2, 85748 Garching, Germany
   \and
   Universitäts-Sternwarte, Fakultät für Physik, Ludwig-Maximilians-Universität München, Scheinerstr. 1, 81679 München, Germany
    }

\abstract
{Supernova remnants (SNRs) and superbubbles (SBs) evolve in complex galactic environments shaped by various different processes. While classical modeling approaches successfully describe blastwaves in uniform media, a unified framework capable of treating these environmental effects simultaneously has been lacking.}
{We develop a flexible analytical framework for modeling blastwave evolution in arbitrary environments and use it to investigate how large-scale galactic structure affects the dynamics and morphology of evolved SNRs and SBs.}
{We introduce FRANZ (FRamework for ANalytical one-Zone blastwave dynamics), a modular thin-shell model that follows the local evolution of a shock-surface segment in environments characterized by arbitrary density, velocity and gravitational fields. After validating the model against well-established analytical results, we apply it to study the effects of vertical stratification, galactic shear and dense galactic substructure on blastwave evolution.}
{FRANZ reproduces the classical evolution of blastwaves in uniform media while extending to complex environments. We derive criteria for disk break out in stratified media, characterize the timescales on which differential rotation deforms blastwaves and identify a new mechanism, by which it can suppress the momentum coupling in continuously-driven blastwaves. 
Interactions with dense filaments modify both shock-surface morphology and dynamics and confound the interpretation of the expansion history of observed remnants, which depends on the density distribution prior to the onset of explosions, which is fundamentally inaccessible from the observed state. In highly structured media with a high volume-filling factor of diffuse gas, ages inferred from the observed state may be systematically overestimated.}
{FRANZ provides a computationally inexpensive and extensible framework for studying blastwaves in realistic galactic environments. It offers a useful complement to numerical simulations for interpreting observations of evolved SNRs and SBs and for developing improved models of stellar feedback in large-scale simulations.}


   \maketitle

\section{Introduction} \label{sec:intro}

The environment in which supernovae (SNe) explode is neither stationary nor homogeneous. Observations of Galactic supernova remnants (SNRs) reveal complex characteristics that are commonly attributed to interactions with structured circumstellar \citep[e.g.,][]{2024ApJ...961...32K, 2024ApJ...976L...4D} and interstellar media \citep[e.g.][]{2018A&A...612A.110A, 2023MNRAS.518.2320D, 2024MNRAS.52711685P}, while extragalactic observations indicate that large-scale gas flows and gravity can further influence their evolution \citep{2023ApJ...944L..24W}.

Although numerical studies can reproduce the properties of individual remnants, such as G1.9+0.3 \citep{2023ApJ...942...94Z}, G332.5-5.6, G290.1-0.8 \citep{2023MNRAS.519.5358V}, Pa 30 \citep{2024arXiv240313641D}, and the Local Bubble \citep{2006A&A...452L...1B, 2016Natur.532...73B}, analytical models generally assume expansion into uniform, stationary media \citep[e.g.][]{1988RvMP...60....1O, 1999ApJS..120..299T, 2019MNRAS.490.1961E}. Only a limited number of studies have explored departures from this idealization \citep[e.g.,][]{1969JFM....35...53L, 2016MNRAS.460.2962H, 2024ApJ...960...81J, 2025MNRAS.540.1124L}.

These analytical models provide a well-established evolutionary picture in which SNRs progress from free expansion\citep{1999ApJS..120..299T} through the Sedov-Taylor phase\citep{1950RSPSA.201..159T, 1959sdmm.book.....S} before radiative losses lead to shell formation and subsequent snowplow evolution\citep{1988ApJ...334..252C, 2015ApJ...802...99K, 2016MNRAS.456..710F}, ultimately merging with the ISM\citep{1992ApJ...392..131S, 2024ApJ...965..168R} or contributing to the formation of superbubbles \citep[SBs; e.g.,][]{2019MNRAS.490.1961E, 2022ApJS..262....9O}. While this framework successfully describes idealized environments, its applicability to structured galactic media shaped by turbulence, differential rotation and gravity remains uncertain.

Previous analytical studies have typically considered individual environmental effects in isolation. Vertical stratification has been shown to accelerate adiabatic blastwaves after atmospheric breakout, although these calculations generally neglect radiative cooling and gravity, which become important on comparable timescales \citep{1960SPhD....5...46K, 1969JFM....35...53L, 1976A&A....50..105M, 1990ApJ...354..513K}. Differential rotation deforms SNRs into elongated morphologies over a fraction of an orbital period and may explain the shapes of observed superbubbles \citep{1987A&A...186..287T, 1995RvMP...67..661B, 2020A&A...644A..72P, 2024ApJ...960...81J}.

The effects of turbulent density fluctuations and small-scale low-density channels have likewise been investigated \citep{2016MNRAS.460.2962H, 2025MNRAS.540.1124L}. These studies suggest that turbulence only weakly modifies the average radial momentum of SNRs, whereas localized channels can substantially alter the coupling of SN feedback to the surrounding ISM. A unified analytical framework that incorporates these mechanisms simultaneously is, however, still lacking.

In a recent paper \citep[Henceforth RBB25][]{2025A&A...702A..12R} we introduced the SISSI (Supernovae In a Stratified, Shearing ISM) simulations, in which we study the evolution of SNRs in a realistic galactic environment.
There we have found that, while the dynamics of young and small SNRs are well described by the simple analytic models based on SNe exploding in a uniform, stationary ISM, once they reach a certain age ($\gtrsim 1 \, \text{Myr}$) and size ($\gtrsim 100 \, \text{pc}$) SNRs begin to deviate from the classical theory, likely due to the above mentioned processes. 
Thus, in order to better understand the dynamics of such large and old SNRs, a comprehensive model that takes all of these effects into account is needed.

In this paper we introduce a FRamework for ANalytical one-Zone blastwave dynamics (FRANZ) in order to model the the dynamics of SNRs in complex environments, taking into account the findings of \citetalias{2025A&A...702A..12R}.
The model aims to lay out the theoretical foundations for the study of SNRs in complex geometries and serve as a simple tool for exploring the effects of previously unexplored phenomena on SNR dynamics, without the need of computationally challenging and expensive numerical simulations. 
To achieve this goal we aimed to formulate the model in a modular way, that easily allows for the inclusion of new phenomena.  

The remainder of this paper is organized as follows. In Sec.~\ref{sec:model} we describe our model. In Sec.~\ref{sec:model_application} we apply the model to a number of complex settings, in order to investigate their individual effects on SNR evolution. In Sec.\ref{sec:discussion} we discuss some of the limitations and future directions of the model. We close in Sec.~\ref{sec:summary} summarizing our results.
In the Appendix we provide some additional background to some aspects of our model and evaluate how well it reproduces a number of well established results.

\section{The FRANZ model}\label{sec:model}

\begin{figure*}
\centering
 \includegraphics[width=0.8\linewidth, clip=true]{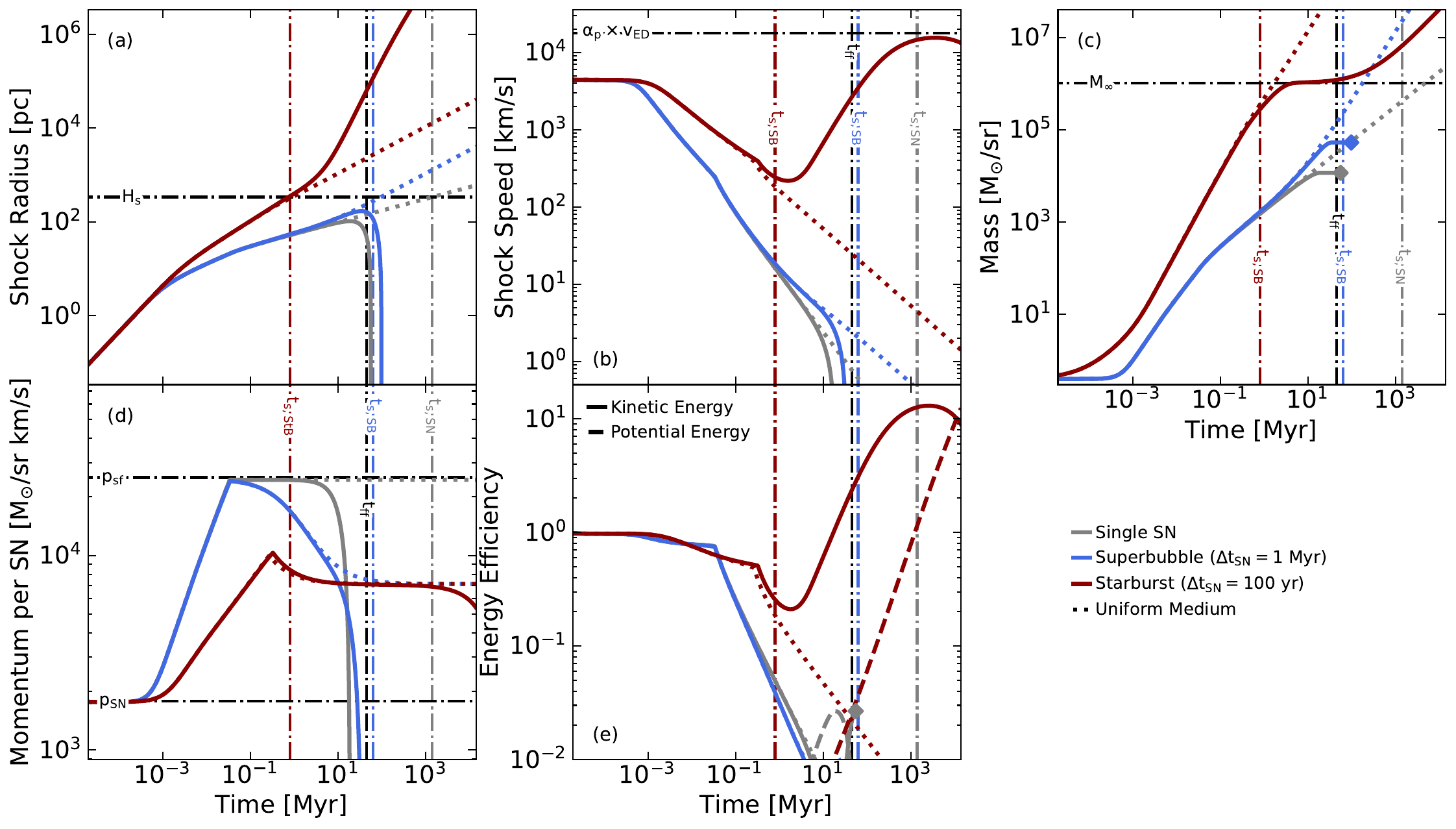}
 \caption{Time evolution of shock radius (a), shock speed (b), swept-up mass (c), momentum per SN (d), and energy efficiency (e) in the case of blastwaves expanding into a vertically stratified atmosphere.
 For comparison, dotted lines corresponding to the same models expanding into a uniform medium are shown. 
 Dash-dotted lines depict various characteristic scales.
 Both the single SN and the SB stall without breaking out of the galactic disk within about a free-fall timescale, due to the effect of gravity. 
 Only the starburst can resist the gravitational field of the disk and drive a galactic wind.} 
 \label{fig:stratification}
\end{figure*}

\begin{figure*}
\centering
 \includegraphics[width=0.8\linewidth, clip=true]
 {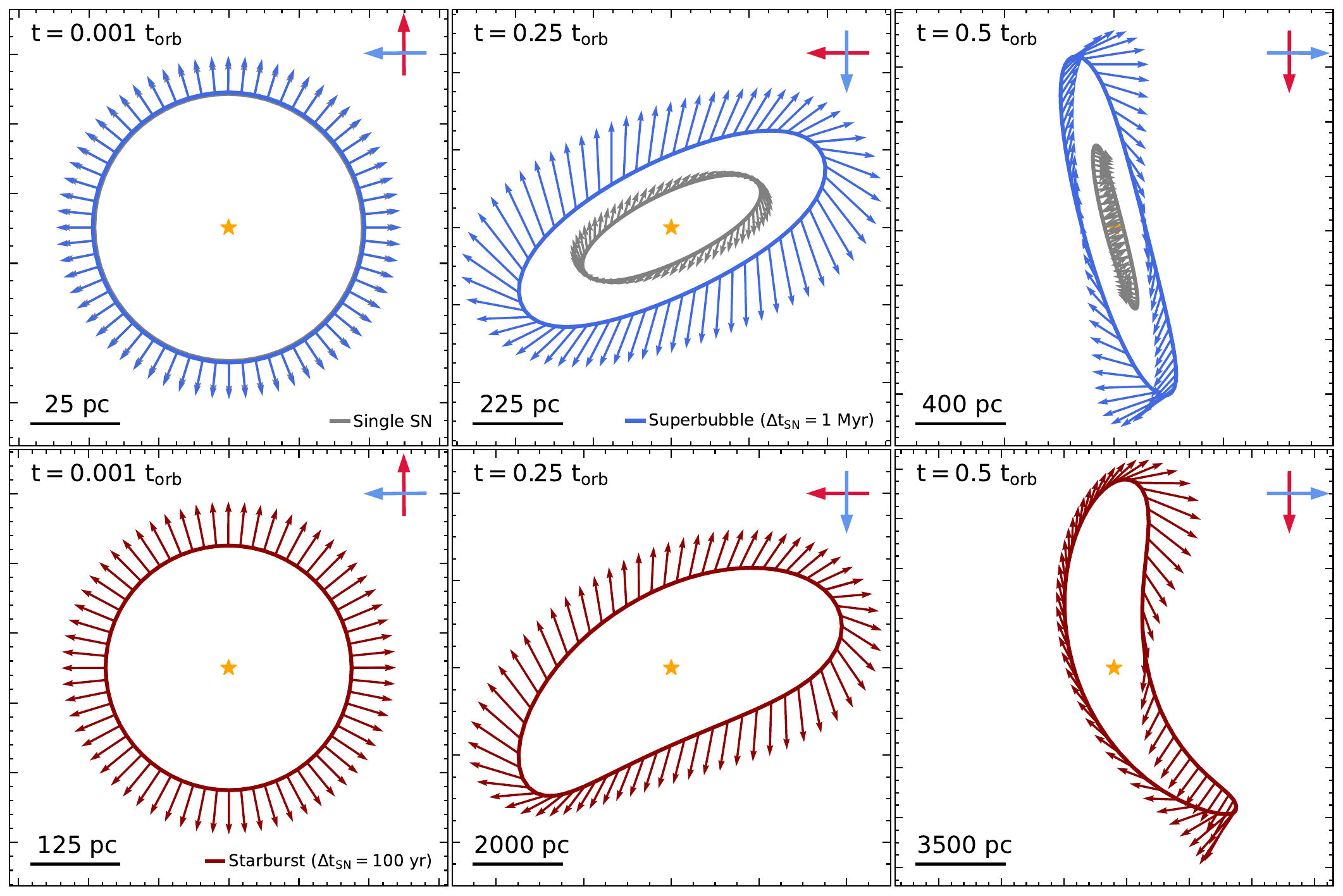}
 \caption{Slices through the xy-plane of the shock-surface of different blastwave models expanding in a shearing, uniform density medium ($n_{0} = 1$, $R_{3} = 8$, $V_{\text{rot, }2} = 2$). The velocity vectors on the surface are shown as arrows with arbitrary scaling.
 Top panels show the single SN and SB models; bottom panels show the starburst. Left, center and right panels show slices after $0.1\,\%$, $1/4$ and $1/2$ of an orbit, respectively. Each panel has a compass pointing towards the galactic center (blue) and the direction of rotation (red). While the SNRs are initially spherically symmetric, after $t_{\text{orb}}/4$ they are significantly stretched out with a pitch angle $\sim 30\,^{\circ}$. The velocity vectors clearly show signs of the epicyclic motion described by Eqs. \ref{eq:momentum-oscillation1} - \ref{eq:momentum-oscillation2}.
 After $\sim t_{\text{orb}}/2$ they cease to be star-shaped making it difficult to measure their geometric properties with our methods.} 
 \label{fig:slices_shear}
\end{figure*}

\begin{figure}
\centering
 \includegraphics[width=0.8\linewidth, clip=true]{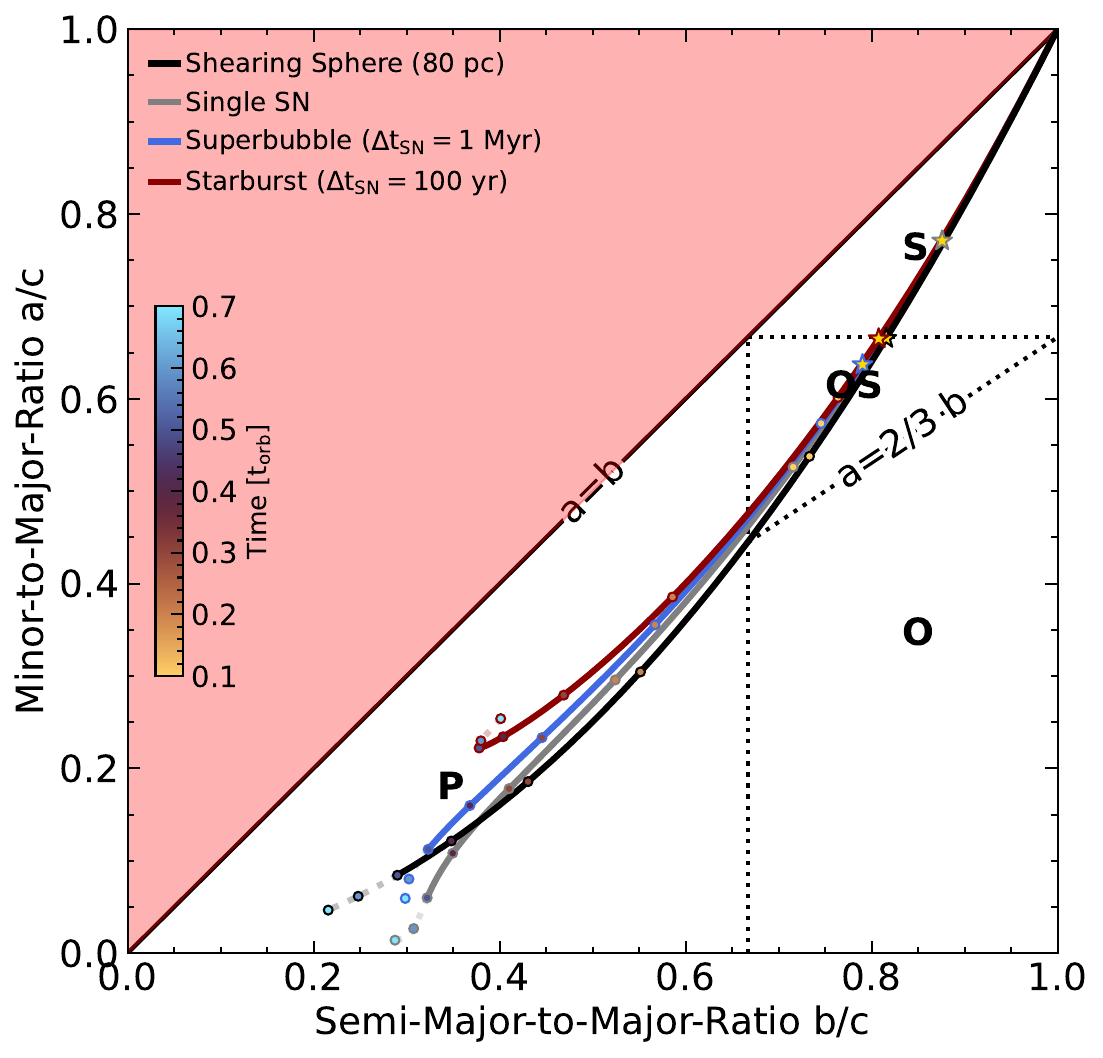}
 \caption{Evolutionary tracks in the shape phase-space of the blastwave models expanding in a shearing, uniform density medium ($n_{0} = 1$, $R_{3} = 8$, $V_{\text{rot, }2} = 2$). In different parts of the phase space the SNRs are either spherical (S), oblate spheroids (OS), prolate (P) or oblate (O). The track of a shearing sphere with a constant radius of 80 pc is shown in black.
 The time at which the blastwaves are expected to cross the $a/c = 2/3$-line is shown as star markers.
 We also plot circle-markers color-coded with the time to provide a reference how long it takes to reach a given degree of deformation. 
 The blastwaves start out as perfect spheres and become increasingly prolate over time. The blastwave models roughly follow the shearing sphere's track, with a slight tendency of more powerful blastwaves towards larger $a/b$.
 After $t_{\text{orb}}/2$ the measured lengths become increasingly unreliable and are thus shown as transparent, dotted lines.} 
 \label{fig:Geometry_Tracks_shear}
\end{figure}

\begin{figure}
 \centering
 \includegraphics[width=0.8\linewidth, clip=true]{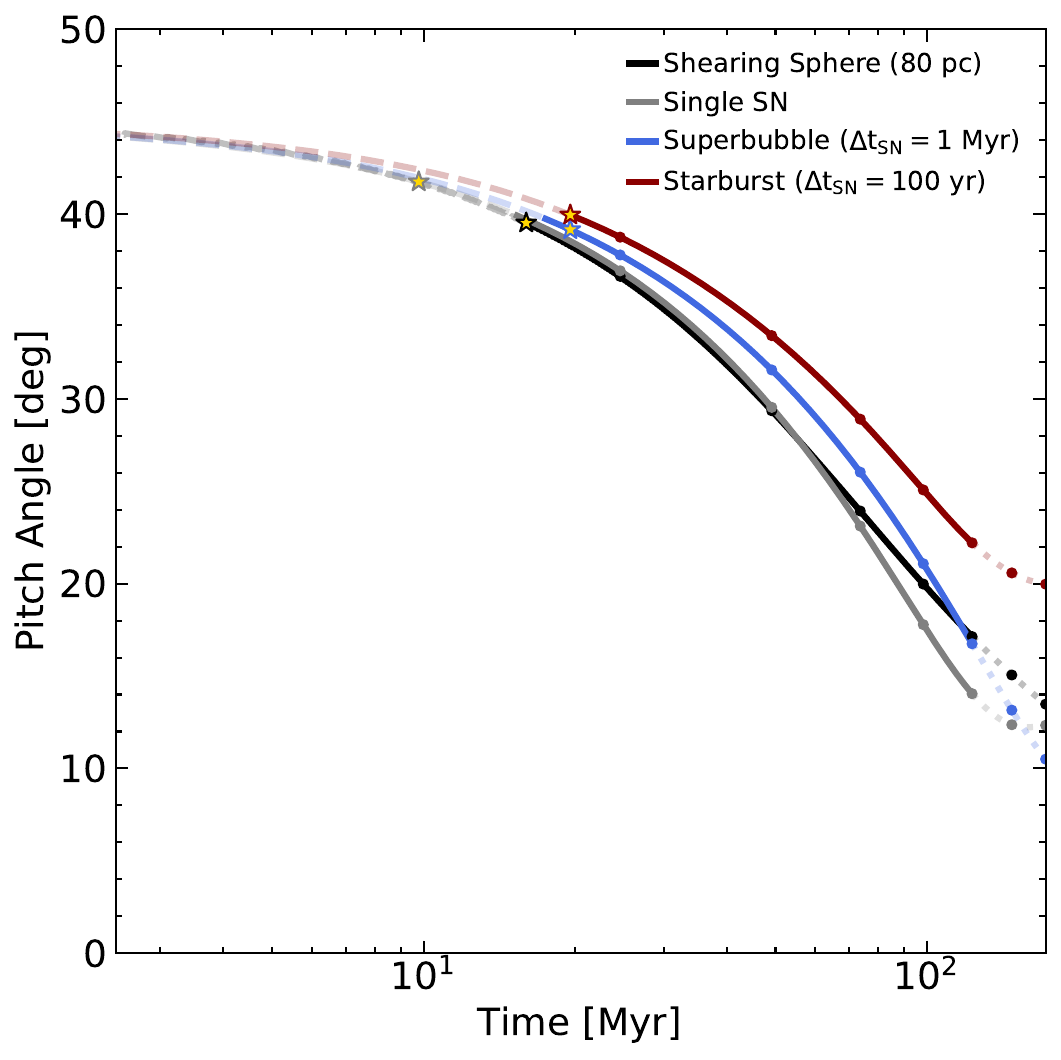}
 \caption{Time evolution of the pitch angle of the major axis of the blastwave models expanding in a shearing, uniform density medium ($n_{0} = 1$, $R_{3} = 8$, $V_{\text{rot, }2} = 2$). For reference the pitch-angle of a shearing sphere with a constant radius of 80 pc is shown in black.
The time at which the blastwaves are expected to cross the $a/c = 2/3$-line is shown as star markers.
During the spherical phase ($a/c > 2/3$) the pitch angle is not very meaningful and is thus shown as transparent dashed lines. After $t_{\text{orb}}/2$ the measured angles become increasingly unreliable and are thus shown as transparent, dotted lines.
The pitch angle starts off near $45\,^{\circ}$ and decays over time. 
The blastwaves roughly follow the pitch-angle evolution of the shearing sphere with a tendency towards larger pitch angles for more rapidly expanding models.} 
 \label{fig:Pitch_angle_shear}
\end{figure}

We introduce the FRANZ\footnote{The FRANZ code is made publicly available at \href{https://zenodo.org/records/21180994}{https://zenodo.org/records/21180994}.} model for the expansion of a blastwave powered by central energy- and mass-injection.
The blastwave is expanding into an arbitrary medium characterized by density-, velocity- and (gravitational) acceleration fields, denoted $\rho_{0}\left(\vec{r}, t\right)$, $\vec{v}_{\text{ext}}\left(\vec{r}, t\right)$ and $\vec{g}\left(\vec{r}, t\right)$, respectively.
Our model solves the blastwave equation of motion \citep{1988RvMP...60....1O} using the thin-shell and sector approximations \citep{1969JFM....35...53L}. It follows the one-zone dynamics of a shell-segment per unit angle (i.e. along a single streamline) by assuming that all of the swept-up mass along the streamline is incorporated into an infinitesimally thin shell.
Similarly to other models using the thin-shell approximation \citep[e.g.][]{1995RvMP...67..661B, 2020A&A...644A..72P, 2024ApJ...960...81J}, we model the curving of streamlines due to normal accelerations, but instead of requiring to evolve the whole shock-surface simultaneously, we opt for an entirely local approach.

The expanding shell segment originates from the explosion center $\vec{r}_{\text{expl}}$ and is associated with its initial expansion direction $\hat{e} = \left(\text{sin }\theta \, \text{cos }\phi, \text{sin }\theta\,  \text{sin }\phi, \text{cos }\theta\right)$. 
We use a HEALPix tesselation of the unit sphere \citep{2005ApJ...622..759G} to uniformly sample all directions.

The dynamics of the shell-segments traced by their position $\vec{r}_{\text{s}}\left(t\right)$, expansion velocity $\vec{v}_{\text{s}}\left(t\right)$ in the frame comoving with the background, mass $M\left(t\right)$ and energy $E\left(t\right)$ are described by the following set of equations
\begin{eqnarray}
            \dot{M} &\equiv& \dot{M}_{\text{in}} + \dot{M}_{\text{sw}} = \dot{M}_{\text{in}} + \rho_{0} \, \vec{\Sigma} \cdot \vec{v}_{\text{s}} \label{eq:mass_cons.} ~,\\
            \frac{\text{d}}{\text{d}t}\left(M \vec{v}_{s}\right) &=& \Delta P \, \vec{\Sigma} + M \left(\vec{g} - \dot{\vec{v}}_{\text{ext}}\right) + \dot{p}_{\text{in}} \vec{n} \label{eq:EoM} ~,\\
            \dot{E}  &=& \dot{E}_{\text{in}} - \dot{E}_{\text{cool}} + M \left(\vec{g} - \dot{\vec{v}}_{\text{ext}}\right) \cdot \vec{v}_{\text{s}} \label{eq:energy_equation} ~,\\
            \dot{\vec{r}}_{\text{s}} &=& \vec{v}_{s} + \vec{v}_{\text{ext}} \label{eq:shell_position}~,
\end{eqnarray}
where $\dot{M}_{\text{in}}$, $\dot{E}_{\text{in}}$ and $\dot{p}_{\text{in}}$ are the central mass-, energy- and momentum-injection rates, $\dot{M}_{\text{sw}}$ is the incorporation rate of swept-up mass, $\dot{E}_{\text{cool}}$ is the energy dissipation rate due to radiative cooling
$\vec{\Sigma} = \text{d}\vec{A}/\text{d}\Omega$ is the (outward-) oriented surface area, $\vec{n} = \vec{\Sigma} / \left\lVert\vec{\Sigma}\right\rVert$ is the outward-pointing normal vector of the shell-surface and $E = E_{\text{th}} + 0.5\, M\, v_{\text{s}}^2$ denotes the sum of the shell's internal and kinetic energy.

In the sector approximation, volumetric quantities such as energy, mass, and momentum are to be understood per unit solid-angle (i.e., $E = \text{d}E/\text{d}\Omega$, etc.).
For brevity of notation we omit this distinction, except where it could lead to confusion.

\begin{table}[t]
\caption{Constants in Eq. \ref{eq:pressure_gradient} as function of the adiabatic index $\gamma$.}
\label{tab:pressure_gradient}
\centering
\begin{tabular}{l|c c c}
   & $c_{0}$  & $c_{1}$ & $c_{2}$\\\hline
  Value & $\frac{\gamma-1}{2\left(2\gamma - 1\right)}$ & $\frac{3}{2}\left(\gamma+1\right)^2$ & $3+ \frac{4\gamma}{\gamma+1}$\\
  Value ($\gamma = 5/3$) & 1/7 & 32/3 & 11/2 \\
  \hline      
\end{tabular}%

\end{table}

We evaluated the pressure gradient force $\vec{F}_{\Delta P} = \Delta P \, \vec{\Sigma}$ using the formalism of \citet{1969JFM....35...53L} which leads to the expression
\begin{equation}\label{eq:pressure_gradient}
    \Delta P = c_{0} \left[c_{1} \frac{E}{3 \delta V} + 3 \rho_{0} v_{\text{s}}^2 + \left(k_{\rho} - c_{2}\right) \frac{M\, v_{\text{s}}^2}{3 \delta V}\right] ~,
\end{equation}
where we list the constants $c_{i}$ in Tab. \ref{tab:pressure_gradient}, 
\begin{equation}
    k_{\rho} = -\frac{\text{dlog}\,\rho_{0}}{\text{dlog}\,r_{\text{s}}} ~,
\end{equation}
and we substituted $r_{\text{s}}^3 \rightarrow 3\,\delta V$, since in our framework, which can account for a more general geometry, the radius is not formally defined.

We direct the interested reader to the App. \ref{app:pressure-gradient} for a derivation of Eq. \ref{eq:pressure_gradient}.
Radiative blastwaves are usually modeled by setting the adiabatic index $\gamma=1$ \citep{1988RvMP...60....1O}, which leads to $\Delta P  \propto \left(\gamma-1\right) = 0$. We thus set $\Delta P = 0$ once radiative cooling becomes dominant.

The derivation of Eq. \ref{eq:pressure_gradient} relies on a number of assumptions such as a strong shock, expanding along straight streamlines into a stationary ambient medium. 
Moreover, due to the substitution of $r_{\text{s}}^3 / 3 \rightarrow \delta V$, the used expression for the pressure-gradient force is formally ill-defined as it depends explicitly on a gauge choice. Nonetheless, we find that it provides a reasonable approximation, since it is only applied before radiative effects become relevant and in this regime the dynamics are typically dictated by the spherically symmetric ($\delta V \sim r_{\text{s}}^3/3$) expansion of a strong blastwave with $v_{\text{s}} \gg v_{\text{ext}}$.

The central mass-, energy- and momentum-injection rates are linked
\begin{equation}
    \dot{p}_{\text{in}} = \alpha_{\text{p}} \sqrt{2 \dot{M}_{\text{in}} \dot{E}_{\text{in}}} ~,
\end{equation}
where $\alpha_{\text{p}}$ is a boost-factor that accounts for the coupling of the interior of the SNR to the shell.
For adiabatic blastwaves, this coupling is explicitly accounted for with the pressure gradient force Eq. \ref{eq:pressure_gradient} and we can set $\alpha_{\text{p}} = 1$\footnote{To be completely consistent, $\alpha_{\text{p}} = \sqrt{c_{1} / 2\left(c_{2} - k_{\rho}\right)} \sim 1$, which ensures $\dot{E}_{\text{kin}} = \dot{M}v_{\text{s}}^2/2$ during the ejecta dominated phase.}, 
however \citet{2024ApJ...970...18L} have shown that radiative, continuously powered blastwaves approach a so-called rapidly-cooling wind solution, where the momentum injected into the shell is mediated by the hot interior, leading to slightly boosted momentum-injection with $\alpha_{\text{p}} \sim 4$. 

The energy dissipation rate due to radiative cooling is
\begin{equation}
    \dot{E}_{\text{cool}} = \chi \frac{\Lambda}{\mu^2} \rho_{0} M ~,
\end{equation}
where $\chi = \left(\gamma+1\right)/\left(\gamma-1\right)$ is the shock-compression ratio in the case of a strong shock and $\mu = 1.4\, m_{\text{H}}$ is the mean atomic weight.
We followed \citet{2022ApJS..262....9O} in approximating the cooling rate of an adiabatic blastwave as $\Lambda = 10^{-22}\,\Lambda_{6, -22}\, T_{\text{s, }6}^{-0.7} \,\text{erg s}^{-1}\,\text{cm}^3$, where $T_{\text{s}} = 10^6\, T_{\text{s, }6}\,\text{K} = \tau \mu v_{\text{s}}^2/k_{\text{B}}$ is the post-shock temperature and $\tau = 2 \left(\gamma-1 \right)\left(\gamma+1\right)^{-2}$.

We transitioned to the radiative stage by switching off cooling and pressure gradient forces and setting $\alpha_{\text{p}} = 4$ once $\gtrsim 10\,\%$ of the thus far injected energy have been radiated away, in line with the result that during the radiative transition about half of the thermal energy ($\sim 0.24 \, E_{0}$, see Figure~\ref{fig:notable_limits}) is lost to radiation \citep{1974ApJ...188..501C, 1981MNRAS.195.1011F}.
Raising the threshold value leads to a slightly (up to $\lesssim 25 \, \%$) greater terminal momentum with otherwise marginal effects, as is shown in the Appendix~\ref{app:cooling_threshold}. 

In contrast to previous work \citep[e.g.][]{2020A&A...644A..72P}, we do not modify the dynamics once the shock velocity falls below the velocity dispersion of the ambient medium or the internal pressure falls below the ambient pressure, since simulations do not show any signs of modified dynamics past this point \citep{2024ApJ...965..168R, 2025A&A...702A..12R}.

The surface area and its direction depend on the instantaneous geometry of the shock-surface, which either requires evolving the whole surface or a local parameterization of it in a neighborhood of each point--here we opted for the latter. The surface element is the vector product tangent vectors
\begin{eqnarray}
    \vec{\Sigma}\left(t\right) &=& \vec{\partial}_{\theta} \times \vec{\partial}_{\phi}, \\
    \vec{\partial}_{\theta} &=& \frac{\partial \vec{r}_{\text{s}}}{\partial \theta}, \qquad \vec{\partial}_{\phi} = \frac{1}{\text{sin}\left(\theta\right)} \frac{\partial \vec{r}_{\text{s}}}{\partial \phi} ~,
\end{eqnarray}
which are known at $t = 0$ and where we divided out the constant $\text{sin}\left(\theta\right)$ for regularity near the poles\footnote{This factor only enters in surface-integrals where now $\Delta\Omega = \text{sin}\left(\theta\right) \Delta\theta\Delta\phi$.}.
We describe their dynamical evolution in the Appendix~\ref{app:tangent_vectors}.

Finally, we defined\footnote{There is some freedom in the definition of the local volume element. For instance, by applying the divergence theorem the dot-product of any function $\vec{F}$ with $\text{div}\, \vec{F} = 1$ with $\vec{\Sigma}$ defines a viable volume-element. The equivalence class of volume-elements corresponds to such functions differing only by a total derivative on the shock surface.} the local volume-element as 
\begin{equation}
    \delta V = \frac{1}{3} \left(\vec{r}_{s} - \vec{r}_{\text{expl}}\right) \cdot \vec{\Sigma} ~,
\end{equation}
where we allowed the explosion center $\vec{r}_{\text{expl}}\left(t\right)$ to move as a function of time to model cases where it is not stationary, such as a star cluster following a circular orbit in a galactic disk.

We confirm both analytically and numerically that FRANZ reproduces the well established behavior of blastwaves in uniform media in the Appendix~\ref{app:notable_limits}.

\subsection{Initial conditions}\label{sec:ICs}

FRANZ is general enough to describe a variety of different types of blastwaves, such as SNe, stellar winds and active galactic nuclei, which may require different initial conditions as well as models for their environment.
Here we focused on SN-driven SBs, which begin to expand following a central point-explosion that deposits a large amount of energy $E_{0}$ and mass $M_{0}$ at $t=0$.
Due to the initial spherical symmetry $\vec{\Sigma} = r_{\text{s}}^2 \, \hat{e}$, where $r_{\text{s}}$ is the shock radius. Regularity of the solution at $r_{\text{s}}=0$ requires the terms in $\Delta P\,r_{\text{s}}^2$ that are $\propto r_{\text{s}}^{-1}$ to cancel as $r_{\text{s}} \rightarrow 0$, which leads to $\left(r_{\text{s}}, v, M, E\right) = (0, \sqrt{cE_{0}/M_{0}}, M_{0}, E_{0})$, where $c=c_{1}/\left(c_{2} - k_{\rho}\right)$.

Despite this cancellation, we started our calculation from a slightly advanced state with $r_{\text{s}} \gtrsim 0$ to avoid numerical problems due to division by zero.

\subsection{Breakdown of the model}\label{sec:breakdown}

While FRANZ is applicable under a wide range of conditions, there are regimes in which it ceases to yield physically meaningful results. Below we summarize common causes of breakdown and their physical interpretation.

In models with gravity, the shock expansion may reverse and lead to collapse. In this situation, where $\vec{v}_{\text{s}} \cdot\vec{\Sigma} < 0$, physically we expect the shock to stop sweeping up mass. To prevent a formal breakdown of the model, we thus set $\dot{M}_{\text{sw}} = 0$.
If the shock subsequently re-expands, care must be taken to avoid double-counting material that may have been swept up previously.

Provided the surface normal initially points outward, the topology ensures that it remains outward-facing unless it undergoes a continuous transition to an inward orientation, which requires passing through a singular point with $\left\lVert\vec{\Sigma}\right\rVert = 0$.
Using Eq.~\ref{eq:tangent_evolution}, the time evolution of the surface normal is given by
\begin{equation}
    \frac{\text{d}}{\text{d}t} \vec{\Sigma} = \left(\vec{\nabla} \cdot \dot{\vec{r}}_{\text{s}} - \left(\vec{\nabla}\dot{\vec{r}}_{\text{s}}\right)^{\text{T}}\right) \vec{\Sigma} ~,
\end{equation}
which shows that converging flows within the tangent plane of the shock can drive $\vec{\Sigma}$ to zero. We thus interpret this singularity as the crossing of neighboring streamlines. Beyond this point the local description is no longer valid. Physically, such behavior may correspond to strong compression, or even collapse, of swept-up material and it may therefore be of astrophysical interest to identify where this behavior occurs.

Finally, in certain setups--such as an adiabatic shock expanding into a medium of finite mass--the shock velocity can diverge in finite time \citep{1969JFM....35...53L, 1990ApJ...354..513K}. The appearance of such divergent or otherwise unphysical behavior signals the breakdown of one or more underlying assumptions, such as neglected physics (e.g., radiative cooling or gravity) or an idealized environment, and must be addressed on a case-by-case basis.

\section{Application: SNRs in SISSI} \label{sec:model_application}

We used FRANZ to quantify how a realistic galactic environment can affect the properties of SNRs.
We particularly focused on ages $\gtrsim 1 \,\text{Myr}$, where our numerical simulations have begun to deviate from previous analytical models \citepalias{2025A&A...702A..12R}.
We considered three different explosion models--all with radiative cooling:
\begin{enumerate}
    \item A single SN explosion (SN).
    \item An SB powered by 1 SN every $\Delta t_{\text{SN}} = 1\,\text{Myr}$\ (SB).
    \item A starburst powered by 1 SN every $\Delta t_{\text{SN}} = 100\,\text{yr}$\ (StB).
\end{enumerate}

Each SN injects a total energy of $E = 10^{51}\, E_{51}\,\text{erg}$ and a mass of $M_{\text{ej}} = M_{\text{ej, }0}\,\text{M}_{\odot}$, corresponding to a momentum of $p_{\text{SN}} \sim 10^4\, p_{4}\,\text{M}_{\odot}\,\text{km s}^{-1}$ across the whole sky. For convenience we write $\Delta t_{6} = \Delta t_{\text{SN}} / \text{Myr}$ and $\dot{p}_{\text{in}} = 10^{4} \,\dot{p}_{4} \, \text{M}_{\odot} \, \text{km s}^{-1}\,\text{Myr}^{-1}$.

For the environmental effects that might affect SNR evolution we considered vertical stratification, galactic rotation and the effect of dense substructures.
To obtain an intuition for each, we first considered each effect separately, before discussing their combined effect in concert.

\subsection{Vertical stratification}\label{sec:stratification}

\begin{figure*}
\centering
 \includegraphics[width=0.8\linewidth, clip=true]{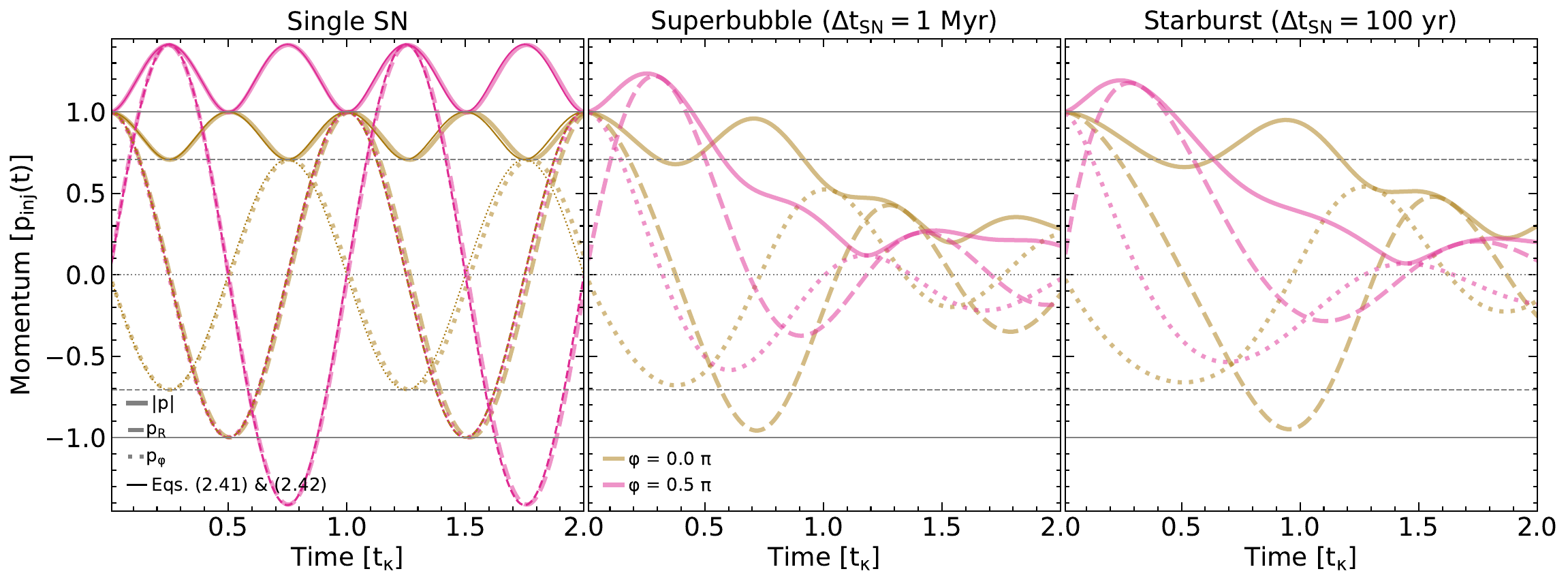}
 \caption{Time evolution of the momentum components of blastwaves expanding in a shearing, uniform density medium ($n_{0} = 1$, $R_{3} = 8$, $V_{\text{rot, }2} = 2$). Different panels correspond to the different explosion models. Solid (dashed, dotted) lines correspond to the magnitude (radial-, azimuthal component) of the momentum vector. In the first panel, thin lines depict the expectation the analytical considerations matching well with the numerical results. Horizontal lines correspond to integer multiples of the radial momentum $p_{\text{inj}}$ and $p_{\text{inj}}/\sqrt{2}$ of a model without shear.} 
 \label{fig:momentum_shear}
\end{figure*}

While the theory of adiabatic blastwaves expanding into a stratified atmosphere is well established, most astrophysical blastwaves are expected to cool long before they can be affected by vertical stratification.
Moreover, blastwaves breaking out of the galactic disk are subject to the galaxy's gravitational pull, which can keep shocks from leaving the galactic ecosystem \citep{2025A&A...701L...5R}.
However, so far these effects have received only little attention \citep{1995RvMP...67..661B, 2022ApJ...932...88O, 2024ApJ...960...81J}.

To address this gap, we applied FRANZ to the vertical expansion of a blastwave from the midplane of an isothermal slab in vertical hydrostatic-equilibrium.
The density profile and gravitational acceleration are given by
\begin{eqnarray}
    \rho_{0}\left(z\right) &=& \rho_{\text{mp}} \, \text{cosh}^{-2}\left(\frac{z}{H_{\text{s}}}\right) \label{eq:density_profile} ~,\\
    g_{z}\left(z\right) &=& - 2 \frac{\sigma^2}{H_{\text{s}}} \text{tanh}\left(\frac{z}{H_{\text{s}}}\right) \label{eq:gravitational_field_strat}~,
\end{eqnarray}
where 
\begin{equation}\label{eq:scale_height}
    H_{\text{s}} = \frac{\sigma}{\sqrt{2\pi G \rho_{\text{mp}}}}  \sim 338 \,\sigma_{1} n_{0}^{-0.5}\, \text{pc}
\end{equation}
is the vertical scale height, $\sigma = 10 \, \sigma_{1} \, \text{km s}^{-1}$ is the velocity dispersion of the ISM and $\rho_{\text{mp}} = \mu \, n_0 \, \text{cm}^{-3}$ is the midplane gas density.
In this setup the mass that can be swept up by a blastwave approaches a finite value
\begin{eqnarray}
    M_{\infty} &=& \int_0^{\infty} \rho_{0}\left(z\right) z^2 \text{d}z \nonumber \\ 
    &=& \frac{\pi^2}{12} \rho_{\text{mp}} H_{\text{s}}^3 \sim 10^{6} \, \sigma_{1}^{3} \, n_{0}^{-1/2} \, \text{M}_{\odot} \ . 
\end{eqnarray}

We expect the dynamics to be strongly affected after about a free-fall timescale
\begin{equation}\label{eq:t_ff}
    t_{\text{ff}} = \sqrt{ \frac{3 \pi}{32 G\rho}} \sim 44.9 \, n_{0}^{-0.5} \, \text{Myr} ~,
\end{equation}
or after the blastwaves break out of the disk after
\begin{eqnarray}
    t_{\text{s}} &\sim&  1.4 \, \sigma_{1}^4 \, n_{0}^{-0.87} \, E_{51}^{-0.93} \, \text{Gyr} ~~~ \left(\Delta t_{\text{SN}} \rightarrow \infty\right),\\
    t_{\text{s}} &\sim&  294 \, \sigma_{1}^2 \, n_{0}^{-0.5} \, \dot{p}_{4}^{-0.5} \, \text{Myr} ~~~~~~ \left(\Delta t_{\text{SN}} \ll t_{\text{s}}\right),
\end{eqnarray}
whichever occurs first.

We show in Figure \ref{fig:stratification} the dynamical evolution of the shock radius, the shock speed, the swept-up mass, the momentum per SN and the kinetic- and potential energy efficiencies per unit injected energy for the part of different blastwave models expanding vertically into a stratified medium with finite mass.
The single SN and the SB model do not manage to break out from the midplane, due its strong gravitational pull, which starts to significantly affect the expansion momentum and speed after $\sim 1/4 \, t_{\text{ff}}$.
The blastwaves stall after $\sim 1/2 \, t_{\text{ff}}$ and $\sim t_{\text{ff}}$, in the SN and the SB cases, respectively, followed by fall-back onto the disk, leading to an increase in kinetic energy by converting the potential energy that was spent trying to escape the galactic potential. 
By contrast, The starburst model is powerful enough to overcome the gravitational potential and drive a galactic wind.
After breaking out of the disk, the mass encountered by the shock is negligible, while the momentum keeps growing at a constant rate, leading to a constant acceleration. However, after $\sim 200 \, \text{Myr}$ the mass injected at the source, becomes comparable to the swept-up mass, and the gravitational force $\propto M$ begins to grow and overpower the constant force of the central starburst.
We note however, that the assumptions of our model cease to be valid on these vast spatial and temporal scales. \citet{2025A&A...701L...5R} discuss the case of the starburst and its limitations in greater detail.

These results suggest that a blastwave will break out of the disk if $t_{\text{s}} \ll t_{\text{ff}}$, which leads to the condition
\begin{eqnarray}\label{eq:break-out1}
    E_{51} &\gg& 40 \, \sigma_{1}^{4.3}\,n_{0}^{-0.4} ~~~ \left(\Delta t_{\text{SN}} \rightarrow \infty\right),\\
    \frac{p_{4}}{\Delta t_{6}} &\gg& 43 \,\sigma_{1}^4 ~~~~~~~~~~~~~~~ \left(\Delta t_{\text{SN}} \ll t_{\text{s}}\right). \label{eq:break-out2}
\end{eqnarray}
However, as shown by \citet{2025A&A...701L...5R} even if a blastwave breaks out of the disk, unless it is powerful enough to escape the galaxy's gravitational potential it will eventually fall back onto the disk. This might be interpreted as fountain flow.

These results highlight the importance of radiative cooling and gravity for the dynamics of blastwaves in stratified media. While we have neglected the change in the surface area element in our analytical derivation, we confirm that the deviations from $\left\lVert\vec{\Sigma}\right\rVert = r_{s}^2$ are negligible before the onset of collapse. Moreover, these results support the finding of \citetalias{2025A&A...702A..12R} that the minor axis of the simulated blastwaves tends to be aligned with the galactic polar / vertical direction, especially for the SNRs in the densest regions, where the free-fall timescale is more comparable to the simulated $10 \, \text{Myr}$.

We note however, that the simplistic picture of a uniform galactic midplane in vertical hydrostatic equilibrium lacks potentially relevant details, such as low-density channels carved out by turbulence and previous generations of feedback, through which weak SBs and even single SNe might be able to break out of the disk and contribute to the galactic fountain flow or galactic winds.

\subsection{Galactic rotation}\label{sec:shear}

Observational evidence \citep{2023ApJ...944L..24W} and theoretical studies \citep{1995RvMP...67..661B, 2024ApJ...960...81J} suggest that large SBs may be subject to galactic shear, which can stretch out their geometry along the rotation direction \citep{2020A&A...644A..72P}.
In the appendix of \citetalias{2025A&A...702A..12R} we have presented a simple model for the deformation by galactic shear for the case of the so-called shearing sphere, a sphere of initially constant radius.
According to this model, the volume of the structure is unaffected by the deformation, which becomes significant after $\sim 5-6\,\%$ of an orbit, corresponding to a time of
\begin{equation}
    t_{\text{deform}} \sim 0.065 \, t_{\text{orb}} \sim 4 \, R_{3} \, V_{\text{rot, } 2}^{-1}\,\text{Myr} ~,
\end{equation}
similar to the timescale on which SNRs evolve. 
Here, $R = R_{3}\,\text{kpc}$ is the galactocentric radius, $V_{\text{rot}} = 100\, V_{\text{rot, } 2} \,\text{km s}^{-1}$ is the galactic rotation speed and $t_{\text{orb}} = 2\pi \,R/V_{\text{rot}}$ is the orbital timescale.

However, in contrast to the deformation of a structure of fixed size, in the case of SNRs, the expanding motion may couple to the galactic rotation and affect the dynamics in non-trivial ways, that we explore in this section.

We consider blastwaves in a uniform density medium, subject to differential rotation of the form
\begin{eqnarray}
    \vec{v}_{\text{ext}} = V_{\text{rot}} \hat{\vec{e}}_{\varphi}\left(\varphi\right) ~,\\
    \vec{g} = -\frac{V_{\text{rot}}^2}{R}\,\hat{\vec{e}}_{R}\left(\varphi\right) ~,
\end{eqnarray}
where $\hat{\vec{e}}_{\varphi}$ and $\hat{\vec{e}}_{R}$ are the local unit vectors pointing in the (galactic) azimuthal, and radial direction, respectively, and $\varphi$ is the azimuthal angle. 
As the shell expands, its azimuthal angle changes, leading to an implicit time-dependence of $\vec{v}_{\text{ext}}$ that enters into the equation-of-motion Eq. \ref{eq:EoM} through
\begin{equation}
    \dot{\vec{v}}_{\text{ext}} = \frac{\text{d}\vec{v}_{\text{ext}}}{\text{d}\varphi} \dot{\varphi} = -\frac{V_{\text{rot}}\left(V_{\text{rot}} + v_{\text{s, }\varphi}\right)}{R} \, \hat{\vec{e}}_{R}\left(\varphi\right) ~,
\end{equation}
where $v_{\text{s, }\varphi}$ is the azimuthal component of the expansion velocity. For numerical evaluations we consider blastwaves exploding at $R_{3} = 8$, in a medium with an ambient density of $n_{0} = 1$ and a rotation speed of $V_{\text{rot, }2} = 2$.

In order to estimate the conditions under which SNRs are strongly affected by shear we consider its effect on the blastwave equation of motion Eq. \ref{eq:EoM}. 
Since the timescale for shear to affect the dynamics of an SNR is usually much longer than the shell-formation timescale, here we consider the dynamics of a radiative blastwave under the influence of galactic shear.
The equation of motions for the azimuthal and radial momentum components can be written as
\begin{eqnarray}
    \dot{v}_{\text{s, }\varphi} &=& - \frac{\dot{M}}{M} v_{\text{s, }\varphi} - \frac{V_{\text{rot}}}{R} v_{\text{s, }R} - \frac{v_{\text{s, }\varphi}v_{\text{s, }R}}{R} + \frac{\dot{p}_{\text{in, }\varphi}}{M} ~,\\
    \dot{v}_{\text{s, }R} &=& - \frac{\dot{M}}{M} v_{\text{s, }R} + 2 \frac{V_{\text{rot}}}{R} v_{\text{s, }\varphi} + \frac{v_{\text{s, }\varphi}^2}{R} + \frac{\dot{p}_{\text{in, }R}}{M} ~,
\end{eqnarray}
where subscripts $R$ and $\varphi$ correspond to the (galactic) radial and azimuthal components, respectively.

Since the inertial term $\propto \dot{M}/M$ is $\propto v_{\text{s}}^2/r_{\text{s}}$ it decays faster than the terms $\propto V_{\text{rot}} v_{\text{s}}/R$ as the expansion slows down.
Thus, we expect galactic rotation to become dynamically important when the two terms become comparable, i.e. once $v_{\text{s}} / r_{\text{s}} \sim V_{\text{rot}} / R$ at
\begin{eqnarray}\label{eq:shear_time1}
    t_{\text{shear}} &\sim& \left(8\pi\right)^{-1} t_{\text{orb}} ~~~~~~~~ \left(\Delta t_{\text{SN}} \rightarrow \infty\right),\\
    t_{\text{shear}} &\sim& \left(4\pi\right)^{-1} t_{\text{orb}} ~~~~ \left(\Delta t_{\text{SN}} \ll t_{\text{shear}}\right), \label{eq:shear_time2}
\end{eqnarray}
which are both comparable to the deformation timescale of the shearing-sphere.

We illustrate the SNRs' morphological evolution through slices in the $xy$-plane shown in Fig.~\ref{fig:slices_shear}. Moreover, we provide a quantitative description in Figs.~\ref{fig:Geometry_Tracks_shear} and \ref{fig:Pitch_angle_shear} showing the trajectories in the shape phase-space and the time evolution of the pitch angle, respectively. The shape phase-space tracks the minor-to-major ($a/c$) and intermediate-to-major ($b/c$) axis ratios, while the pitch angle measures the orientation of the shock surface relative to the galactic rotation vector, with $90^{\circ}$ pointing toward the galactic center and $-90^{\circ}$ toward the anti-center. We quantified the geometry of the blastwaves using the same ellipsoidal approximation outlined in \citetalias{2025A&A...702A..12R} (see also App. \ref{app:geometry}).

All SNRs evolve from an initially spherical shape towards increasingly prolate geometries, reaching pitch angles on the order of $\sim 30 ^{\circ}$ by $t_{\text{orb}}/4$. In the single SN and SB models the geometry is quite similar suggesting that shear-induced motion dominates their evolution. By contrast, the starburst develops a more strongly curved geometry (higher $a/b$). In all cases the geometry is well approximated by the shearing sphere, though more powerful SNRs have a tendency towards higher $a/b$ and a slower decline of the pitch angle.
After $\sim t_{\text{orb}}/2$ the SNRs cease to be star-shaped, at which point the ellipsoidal approximation used to characterize their geometry becomes increasingly unreliable, as indicated by kinks in the shape phase-space trajectories.

Curiously, the velocity field exhibits a differential offset from pure expansion, resembling epicyclic motion.
To understand the origin of this motion in the case of the single SN, we can drop the subdominant term $\propto v_{\text{s}}^2 / R$ (assuming $r_{\text{s}} \ll R$) and assume $R \sim const.$ to find a coupled harmonic (epicyclic) oscillation of the momenta, with an oscillation frequency of
\begin{equation}
    \kappa = 2\sqrt{2}\pi \, t_{\text{orb}}^{-1} ~,
\end{equation}
and an associated epicycle timescale $t_{\kappa} = t_{\text{orb}} / \sqrt{2}$.
Due to the factor of $2$ in the radial equation of motion the ratio of the amplitudes of the radial and the azimuthal momentum-oscillations is $p_{\text{s, }R}^{\text{max}} / p_{\text{s, }\varphi}^{\text{max}}= \sqrt{2}$. In the case of consecutive SNe, the momentum injection will lead to deviations from this oscillation. 

By assuming that at the onset of the oscillation, the direction of the motion in the co-rotating frame has hardly changed, we estimate the functional shape of the momentum oscillation by
\begin{eqnarray}
    p_{\text{s}, R}\left(t, \phi\right) &\approx& \sqrt{2}\, p_{\text{s}, \varphi}^{\text{max}}\left(t, \phi\right) \, \text{cos}\left(\kappa t - \phi\right) \label{eq:momentum-oscillation1}~,\\
    p_{\text{s}, \varphi}\left(t, \phi\right) &\approx& - p_{\text{s}, \varphi}^{\text{max}}\left(t, \phi\right) \, \text{sin}\left(\kappa t - \phi\right) ~, \label{eq:momentum-oscillation2}
\end{eqnarray}
where $\phi$ is the initial angle between the motion and the galactocentric radial direction, where the amplitude is given by momentum conservation at the onset of the oscillation
\begin{equation}
    p_{\text{s}, \varphi}^{\text{max}}\left(t, \phi\right) \approx \frac{p_{\text{sf}}}{\sqrt{1 + \text{cos}^2\left(\phi\right)}} ~,
\end{equation}
in particular $p\left(t=0, \phi\right) = \sqrt{p_{\text{s}, R}^2 + p_{\text{s}, \varphi}^2} = p_{\text{sf}}$ in the case of a single SN, while we expect the amplitude to grow accordingly in the case of continuous momentum injection.

We quantify these expectations over two epicycles in Fig.~\ref{fig:momentum_shear}, where for each blastwave model, we track two representative points on the shock surface: one initially moving radially outward and one moving parallel to the galactic rotation.
For a single SN, Eqs.~\ref{eq:momentum-oscillation1} - \ref{eq:momentum-oscillation2} accurately capture the momentum evolution. The neglected term $\propto v^{2}/R$ introduces only a small phase shift that accumulates over multiple epicycles. Galactic shear slightly enhances the momentum in initially azimuthal directions, while it slightly reduces the total momentum in initially radial directions. In contrast, models with consecutive SNe only briefly exhibit oscillatory behavior before the momentum decays. After two epicycles, the continuously driven models retain only $\sim 25\%$ of the injected momentum, indicating that the epicyclic response and the outward-directed injection counteract each other and reduce the overall efficiency of momentum coupling.

\begin{figure*}
\centering
 \includegraphics[width=0.8\linewidth, clip=true]{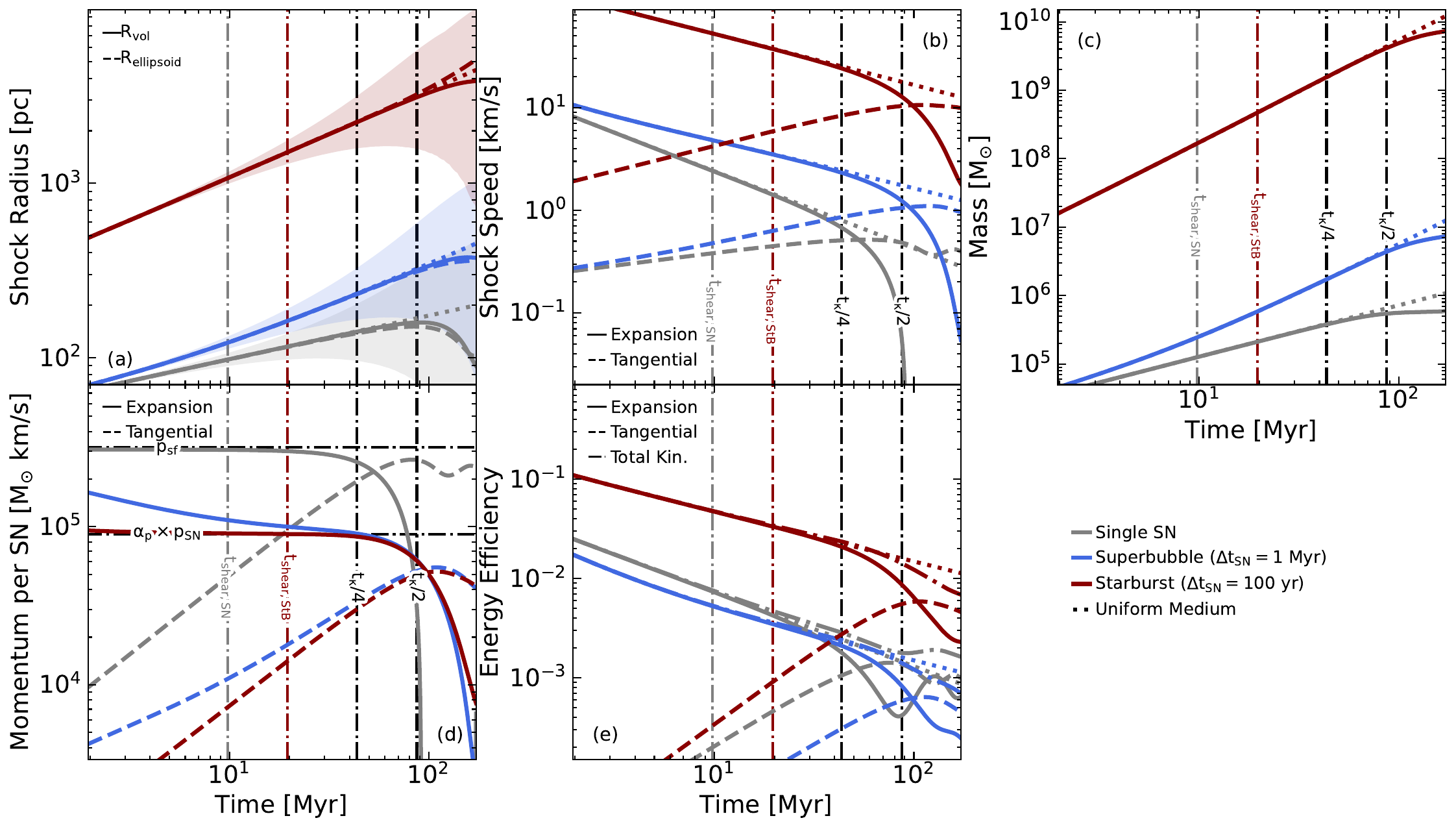}
 \caption{Same as Fig. \ref{fig:stratification} for different blastwave models expanding in a shearing, uniform density medium ($n_{0} = 1$, $R_{3} = 8$, $V_{\text{rot, }2} = 2$). 
 For comparison, dotted lines corresponding to the same models expanding into a uniform medium without shear are shown. 
 Dash-dotted lines depict various characteristic scales.
 In all models the net expansion falls below that of the models without shear from $\sim 80\,\text{Myr}$, despite the epicyclic boost received in certain directions. Instead, a significant amount of the expansion momentum is converted into tangential motion.} 
 \label{fig:globals_shear}
\end{figure*}

We summarize the evolution of the different blastwave models in Fig.~\ref{fig:globals_shear}. Despite transient momentum enhancement in azimuthal directions, we find that galactic rotation suppresses the overall expansion. In all models, after $t_{\kappa}/4 \sim 40\, \text{Myr}$ the velocity starts to fall below that of the models without shear, while from $t_{\kappa} / 2 \sim 80\,\text{Myr}$ onward, the remnant size falls significantly below that of an equivalent model without shear. This matches with the expected decline and reversal of the net expansion speed and radial momentum. At the same time, the tangential momentum grows to values comparable to the initial blastwave momentum, leading to significant tangential motion, comparable to the expansion. In the case of the single SN the total kinetic energy grows larger than in the corresponding model without shear, while in the continuously driven models it falls below.
This mechanism might play an important role for blastwave-driven turbulence generation.

\subsection{Density structures}\label{sec:density_structures}

\begin{figure*}
\centering
 \includegraphics[width=0.8\linewidth, clip=true]{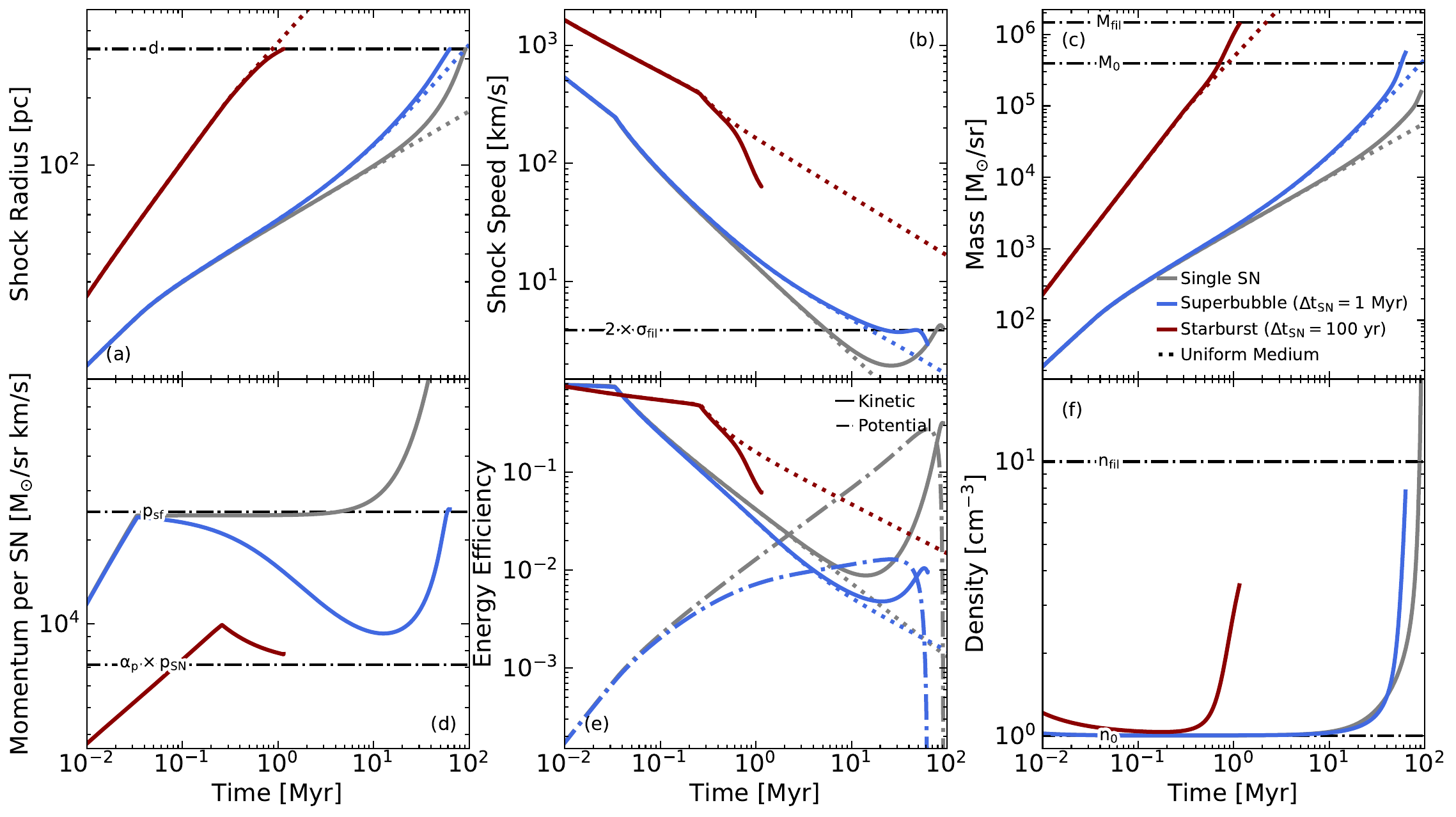}
 \caption{Same as Fig. \ref{fig:stratification} for different blastwave models approaching a filament in a background medium ($n_{\text{fil, }0} = 10$, $T_{\text{fil, }2} = 6$, $d = 8 \, \text{r}_{\text{fil}}$, $n_{0} = 1$). An additional panel (f) shows the average density to account for the environment's multi-phase nature. 
 For comparison, dotted lines corresponding to the same models expanding into the unperturbed background medium are shown. 
 Dash-dotted lines depict various characteristic scales.
 Both the single SN and the SB are trapped in the filament's gravitational potential, while the starburst can overrun it. In all cases the average density increases towards the filament's central density upon approach.} 
 \label{fig:globals_filament_approach}
\end{figure*}

\begin{figure*}
\centering
 \includegraphics[width=0.8\linewidth, clip=true]{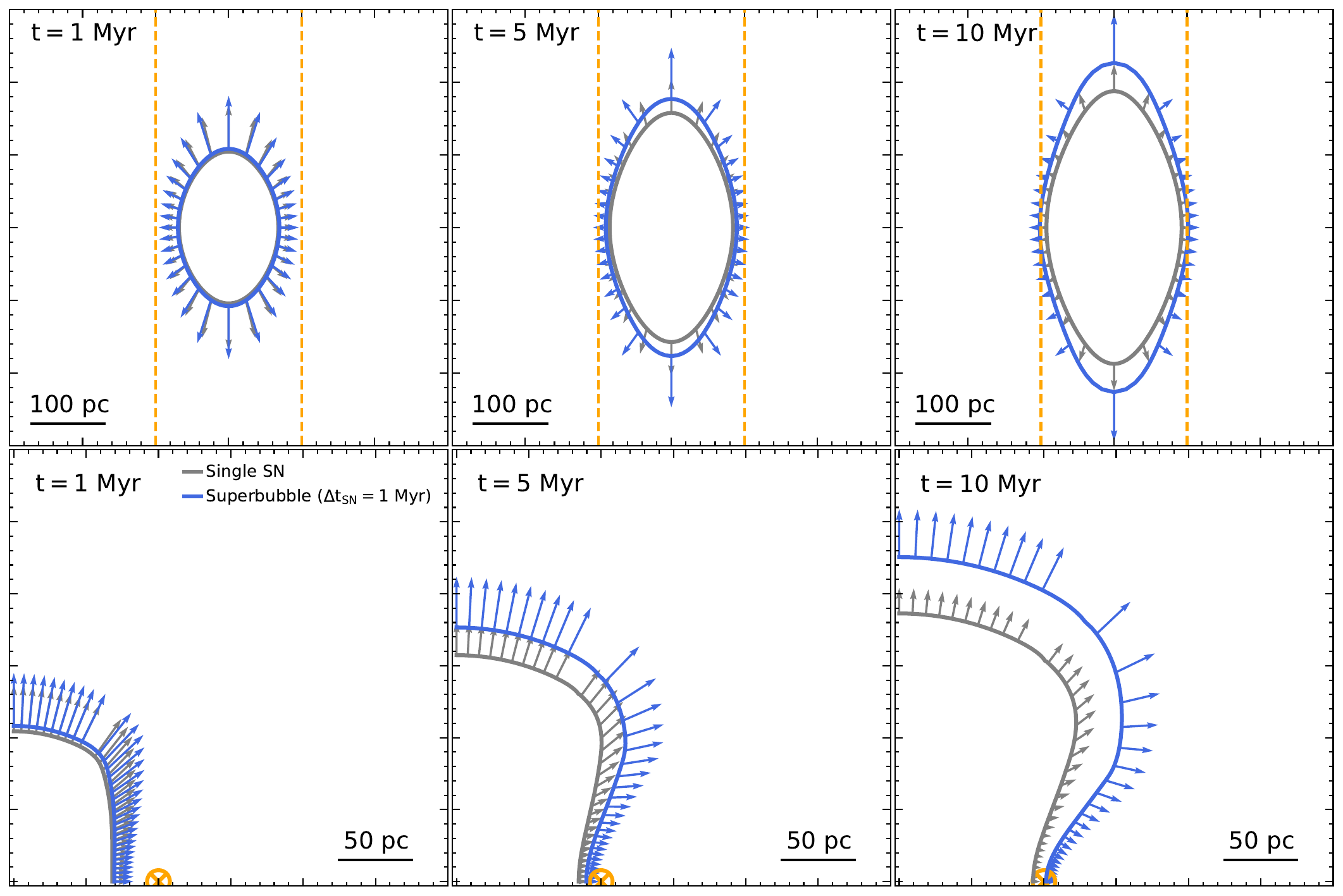}
 \caption{Slices through the shock-surface of different blastwave models expanding into the gap ($n_{\text{gap, }0} = 0.1$) between two filaments 
 ($n_{\text{fil, }0} = 10$, $T_{\text{fil, }2} = 2.65$, $d_{\text{fil}} = 200\,\text{pc}$). The velocity vectors on the surface are shown as arrows with arbitrary scaling.
 Top (bottom) panels show slices through the xy- (yz-) plane. Left, center and right panels show slices after 1, 5 and 10 Myr, respectively. 
 Already at 1 Myr, the SNRs are quite deformed by the geometry of the ISM.
 The ratio between the SNRs' extent parallel to the filaments and towards them grows in time reaching axis ratios $\sim 1/2$ by 5-10 Myr.} 
 \label{fig:slices_filaments}
\end{figure*}

\begin{figure}
\centering
 \includegraphics[width=0.8\linewidth, clip=true]{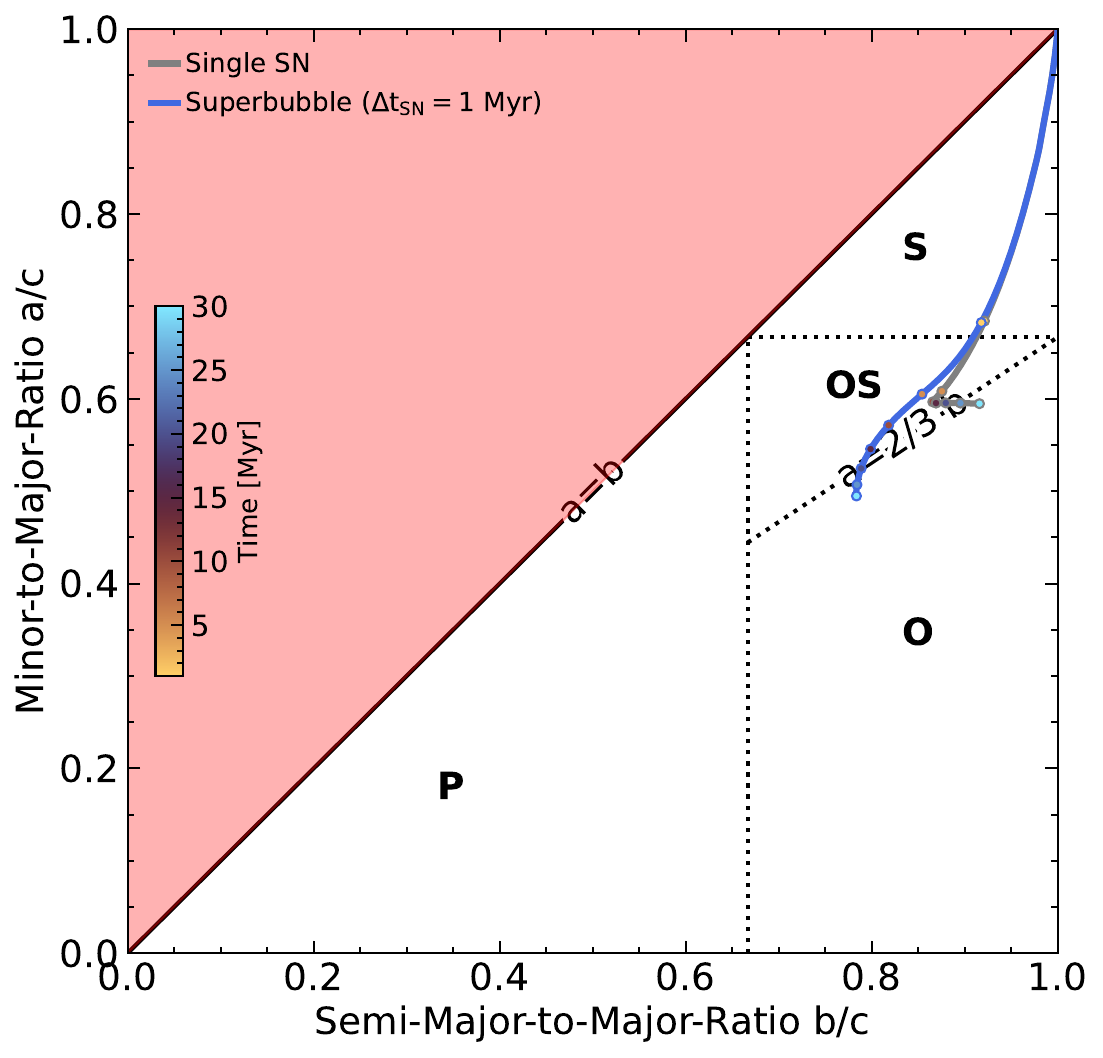}
 \caption{Same as Fig. \ref{fig:Geometry_Tracks_shear} for the different blastwave models 
 expanding into the gap ($n_{\text{gap, }0} = 0.1$) between two filaments 
 ($n_{\text{fil, }0} = 10$, $T_{\text{fil, }2} = 2.65$, $d_{\text{fil}} = 200\,\text{pc}$).
 The blastwaves start out as perfect spheres and becomes increasingly deformed over time.} 
 \label{fig:Geometry_Tracks_filament}
\end{figure}

\begin{figure*}
\centering
 \includegraphics[width=0.8\linewidth, clip=true]{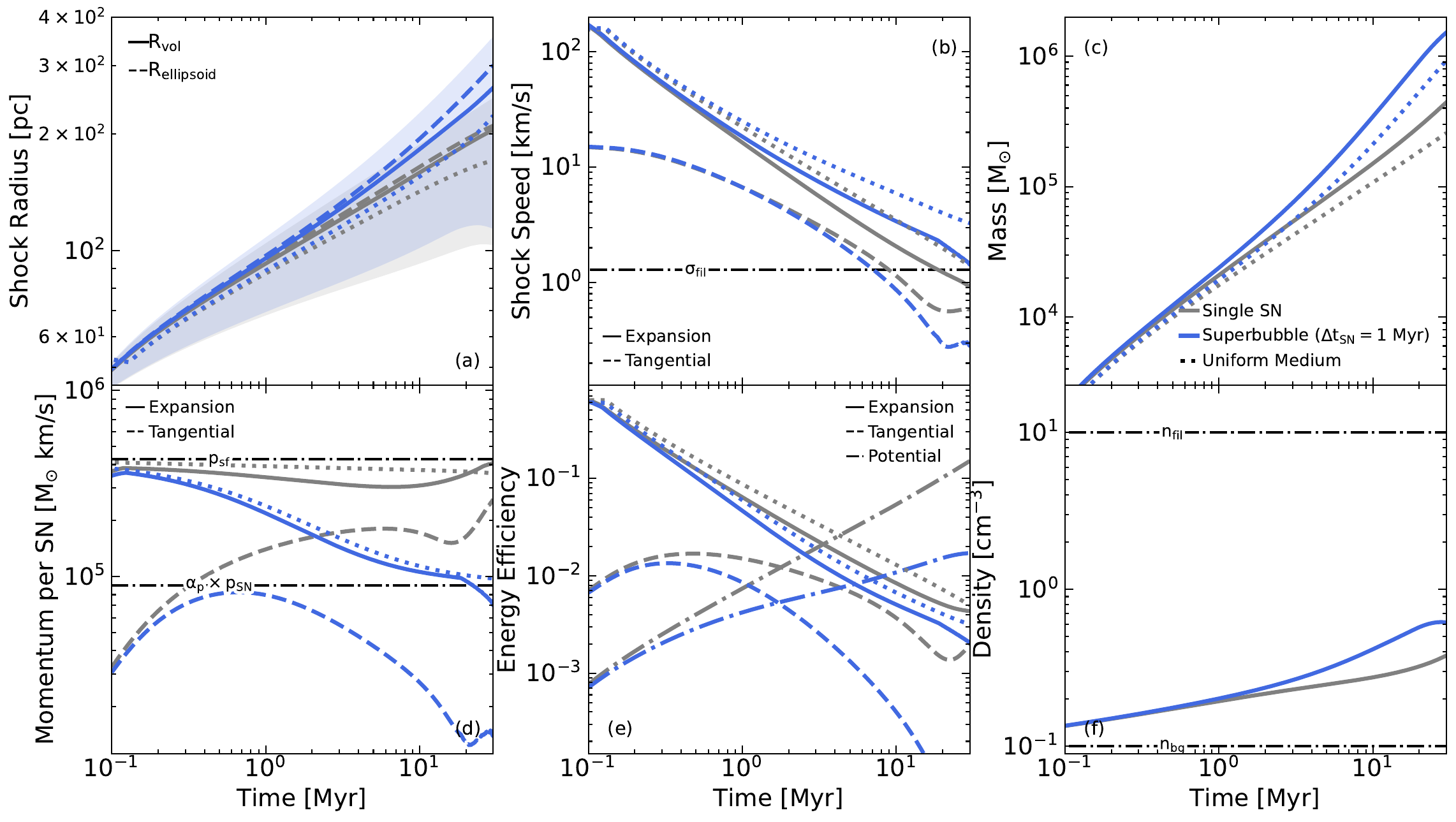}
 \caption{Same as Fig. \ref{fig:globals_filament_approach} for different blastwave models expanding into the gap ($n_{\text{gap, }0} = 0.1$) between two filaments 
 ($n_{\text{fil, }0} = 10$, $T_{\text{fil, }2} = 2.65$, $d_{\text{fil}} = 200\,\text{pc}$).
 For comparison, dotted lines corresponding to the same models expanding into a uniform medium with an ambient density matching the average density of the corresponding inhomogeneous model at each point in time. 
 Dash-dotted lines depict various characteristic scales.
 At the same average density, the blastwaves in inhomogeneous media grow larger than those in homogeneous media as indicated both by the radius panel (a) and the mass panel (c).
 } 
 \label{fig:globals_filament}
\end{figure*}

\begin{figure}
\centering
 \includegraphics[width=0.8\linewidth, clip=true]{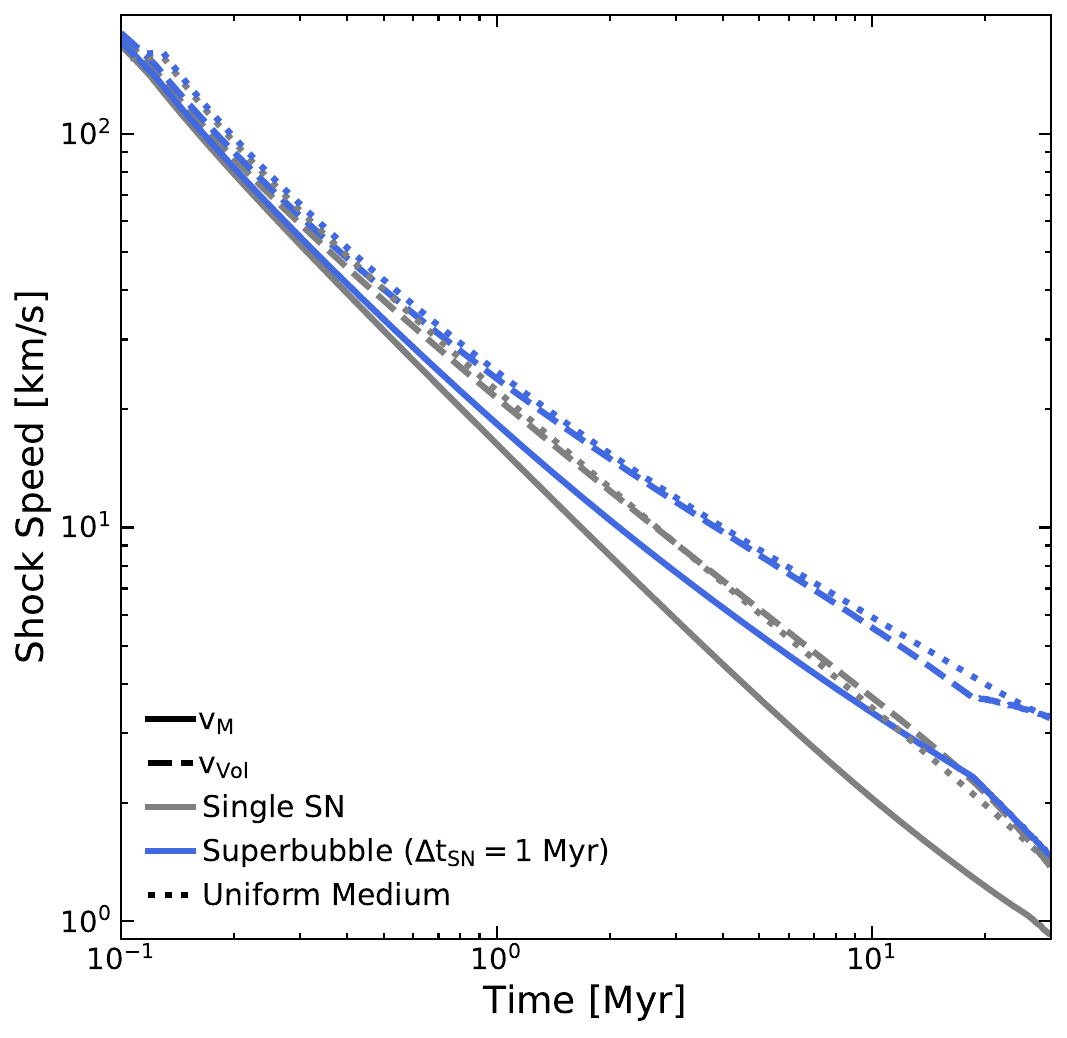}
 \caption{Time evolution of the expansion velocity components of blastwaves expanding into the gap ($n_{\text{gap, }0} = 0.1$) between two filaments 
 ($n_{\text{fil, }0} = 10$, $T_{\text{fil, }2} = 2.65$, $d_{\text{fil}} = 200\,\text{pc}$).
 Solid and dashed lines correspond to the mass-weighted expansion speed and the effective expansion speed obtained by computing the rate of change of the effective radius, respectively
 For comparison, dotted lines corresponding to the same models expanding into a uniform medium with an ambient density matching the average density of the corresponding inhomogeneous model at each point in time. 
 The mass-weighted expansion speed is strongly suppressed due to the stalled expansion in the directions of the filaments, which contribute significantly to the total mass, but little to the total volume. The effective expansion speed roughly matches that of the uniform medium at the time-dependent average density.} 
 \label{fig:expansion_speed_filament}
\end{figure}

Many SNRs are observed to interact with dense structures in their environment \citep{2009ApJ...694L..16H, 2023ApJ...944..110M, 2023ApJ...944L..24W}.
Models and simulations have addressed such interactions of SNRs with their immediate surroundings \citep{2016MNRAS.460.2962H, 2023MNRAS.523.1421M, 2025MNRAS.540.1124L}.
Yet, models addressing the interaction of large SNRs with galactic scale structures, such as molecular filaments, spiral arms and massive clumps remain scarce.

We expect the sizes and separations of such structures to be on the order of the Jeans- and Toomre-lengths~\citep{1902RSPTA.199....1J, 1964ApJ...139.1217T}, which are approximately equal in a marginally stable disk~\citep{1964ApJ...139.1217T},
\begin{equation}\label{eq:jeans_length}
    \lambda_{\text{J}} = \left(\frac{15 \sigma^2}{4\pi G\rho}\right)^{1/2} \sim 926 \, \sigma_{1}\, n_{0}^{-1/2} \, \text{pc} ~.
\end{equation}
On scales comparable to $\lambda_{\text{J}}$, the gravitational acceleration towards such a structure is of order 
\begin{equation}
    g_{\text{J}} \sim \frac{\sigma^2}{\lambda_{\text{J}}} \sim \frac{\sigma}{t_{\text{ff}}} ~.
\end{equation}
Slowly moving objects, with speeds $v \lesssim \sigma$ -- located within about a Jeans-length of the overdensity -- experience free-fall onto it, reaching an asymptotic velocity of order $\sigma$.

In order to test how interactions with density structures may affect blastwave dynamics, we consider various setups featuring isothermal ``Ostriker'' filaments \citep{1964ApJ...140.1056O} in an otherwise uniform background medium. The density profile and gravitational acceleration field of the filament are given by
\begin{eqnarray}
    \rho_{\text{fil}} &=& \frac{\rho_{\text{fil, }0}}{\left(1 + \left(r / r_{\text{fil}}\right)^2\right)^2} ~, \\
    \vec{g}_{\text{fil}} &=& - 4 \frac{\sigma_{\text{fil}}^2}{r_{\text{fil}}} \frac{\vec{r} / r_{\text{fil}}}{1 + \left(r / r_{\text{fil}}\right)^2} ~,
\end{eqnarray}
where $\vec{r}$ is the distance vector from the filament, $r$ its length and
\begin{equation}
    r_{\text{fil}} = 50 \, T_{\text{fil, }2}^{0.5} \, n_{\text{fil, }0}^{-0.5} \, \text{pc} 
\end{equation}
is the filament's scale length, depending on its central density $\rho_{\text{fil, }0} = \mu\, n_{\text{fil, 0}}\,\text{cm}^{-3}$ and its temperature $T_{\text{fil}} = 100 \, T_{\text{fil, }2}\, \text{K}$.

Evidently, if a blastwave occurs very far away from an overdensity, it will cover only a small solid angle on its surface and hardly affect its evolution, while a blastwave occurring within the overdensity can be effectively treated as expanding in a uniform density medium matching that of the overdensity--as long as it is smaller than the overdensity itself.
We estimate the range, within which the overdensity contributes significantly to the blastwaves' dynamics by considering the ratio of the swept-up mass attributable to the overdensity and the background, respectively:
\begin{equation}
    \frac{M_{0}}{M_{\delta}} \sim \frac{d}{\delta\,\lambda_{\text{J, }\delta}} ~,
\end{equation}
where $\delta = n_{\text{fil}} / n_{\text{bg}}$ is the magnitude of the overdensity, $d$ is the distance from the overdensity and $\lambda_{\text{J, }\delta}$ its size.
The impact of an overdensity in this range becomes immediately apparent, when one considers radiative SNRs, where the expansion speed relative to that of an unperturbed SNR is suppressed by the above mass-ratio.

In order to explore the effect of an overdensity on the blastwaves' expansion, in Fig.~\ref{fig:globals_filament_approach} we show the evolution of blastwaves approaching a filament with a central density and temperature of $n_{\text{fil, }0} = 10$ and $T_{\text{fil, }2} = 6$ embedded in a constant density background with density $n_{0} = 1$. The explosions occur at a distance of $d = 8 \, \text{r}_{\text{fil}} \sim 332 \, \text{pc}$ away from the filament, where its contribution to the density amounts to only $\sim 0.1 \, \%$. We stopped the calculation once the blastwave reaches the filament's center.

As expected the starburst overruns the filament. While its speed drops considerably upon reaching the overdensity, it remains well above the escape speed. By contrast, both the SB and the single SN get trapped in the gravitational potential and begin to free-fall approaching the asymptotic free-fall velocity of $\sim 2\, \sigma_{\text{fil}}$. Compared to models without a filament, these models reach higher momentum and kinetic energy due to the filament's gravitational acceleration. 

In all three models, the average density at the position of the filament exceeds that of the background medium. However, the final density differs between the models. The filament’s gravitational field converges neighboring streamlines, leading to a contracting surface area element. While this in turn leads to a slower mass accretion it also shrinks the volume element, ultimately leading to a higher average density.
In the case of the single SN the convergence of streamlines is so advanced that the density exceeds the filament's central density--a signature of gravitational collapse.

In order to study the net effect of galactic substructure on the geometry of blastwaves, we next consider blastwaves expanding in the gap between two parallel filaments--a setup that may naturally occur in galactic disks due to fragmentation and subsequent deformation by differential rotation. We place the filaments at a distance of 200 pc from each other, with a central density of $n_{\text{fil}, 0} = 10$ and a temperature of $T_{\text{fil}, 2} = 2.65$, chosen such that the density in the gap is $n_{0} \sim 0.1$, corresponding to an overdensity of $\chi_{\text{fil}} = 100$. The filaments are embedded in a uniform background medium with a density of $n_{0} = 0.1$. We omit the starburst as it overruns the filaments before 1 Myr.

We illustrate the filaments' effect on the blastwaves' geometry by showing slices through the xy- and yz-planes in Fig.~\ref{fig:slices_filaments} after 1, 5 and 10~Myr of expansion.
At a background density of $n_{0}\sim 0.1$ the blastwaves are expected to reach the filaments after $\sim 1\,\text{Myr}$ and be significantly deformed after a few Myr. For the single-SN (SB) model, the minor-to-major axis ratio decreases to $\lesssim 2/3$ ($\sim 1/2$) after $5,\mathrm{Myr}$ and evolves to $\gtrsim 1/2$ ($\sim 0.45$) after $10,\mathrm{Myr}$ (see also the geometry tracks in Fig.~\ref{fig:Geometry_Tracks_filament}). The portions of the shock surface expanding out of the midplane facing either filament are expanding in a low-density medium, but are pulled toward them, while the midplane region is stalled by the steep density gradient. This combination produces an increasingly concave, rectangular morphology, illustrating how galactic substructure can strongly shape the late-time geometry of SNRs and SBs. 

This geometry differs from the geometry seen in the SISSI simulations~\citepalias{2025A&A...702A..12R}, where lower values of both $a/c$ and $b/c$ are obtained. These differences might arise from vertical stratification, 
more diversity in the galactic substructure and a deeper gravitational potential due to contributions from stars, all of which were neglected here for simplicity.

We summarize the evolution of the different blastwave models in Fig.~\ref{fig:globals_filament}. As expected the presence of the filaments increases the average density encountered by the SNRs. Despite the higher average density, the SNRs grow larger than they would in a uniform medium of the same average density. Curiously however, the mass-weighted expansion speed, obtained by dividing the momentum--which is only slightly affected by the presence of the filaments--by the mass is lower than the expansion speed in a uniform density medium. Moreover, the free-fall-driven acceleration that we found for the regions approaching the filaments in Fig. \ref{fig:globals_filament_approach} is washed out, when considering the entire SNR. 
The presence of the filaments drives tangential motion, decaying from initially $\sim 10\,\text{km/s}$ down to $\lesssim \sigma_{\text{fil}}$ after $\sim 10\,\text{Myr}$ which significantly contributes to the SNRs' momentum- and energy budget in the case of a single explosion, but is rather negligible for the SB.

In order to make sense of the lower expansion speed despite the larger size, in Fig.~\ref{fig:expansion_speed_filament} we compare the mass-weighted expansion speed $v_{M} = p_{\text{exp}} / M$ to the effective expansion speed, defined as the rate of change of the effective radius $v_{\text{vol}} = \dot{r}_{\text{vol}}$. We find that the effective expansion speed roughly matches the expansion speed of a blastwave expanding in a uniform medium with ambient density matching the average density of the blastwave in an inhomogeneous medium. This motivates us to approximate the effective radius as
\begin{equation}\label{eq:volumetric_radius}
    r_{\text{vol}}\left(t\right) \sim \int_{0}^{t} v_{\text{hom}}\left(t', \left\langle\rho\left(t'\right)\right\rangle\right) \text{d}t' ~,
\end{equation}
as opposed to the usually applied
\begin{equation}
    r_{\text{obs}}\left(t\right) \sim \int_{0}^{t} v_{\text{hom}}\left(t', \left\langle\rho\left(t\right)\right\rangle\right) \text{d}t' ~,
\end{equation}
where the difference lies only in the subtle fact, that the former expression requires information about the history of the average ambient density, while the latter depends only on the current ``observable'' density ($\left\langle\rho\left(t'\right)\right\rangle$ vs. $\left\langle\rho\left(t\right)\right\rangle$).
Since in our setup the density increases with time and a higher density implies a lower speed, this suggests that the ``memory'' of a previously encountered lower density medium automatically leads to a larger size compared to a model that expanded throughout with the higher average density at later times. 
Even for the slight overdensity considered here these effects become substantial after a few Myr. Observationally, it has been found that most SNe explode in diffuse gas~\cite{2019MNRAS.483.4551E}. Thus this effect may provide a plausible explanation for the systematically larger radii at fixed average density reported by \citetalias{2025A&A...702A..12R}, and may have important consequences for observational age estimates of SBs that rely on blastwave-models in uniform, stationary media \citep{2022Natur.601..334Z, 2026A&A...707A..98R}.

\subsection{Galactic environment in concert} \label{sec:environment_in_concert}

The previous sections considered the effects of vertical stratification, galactic rotation and large-scale density structure in isolation. In realistic galaxies, however, these mechanisms operate simultaneously. In a marginally stable disk, many of the characteristic scales are linked through the Toomre parameter $Q_{\rm T}$ \citep{1964ApJ...139.1217T}. The free-fall timescale is comparable to the epicyclic timescale, while the Jeans, Toomre and disk scale heights satisfy $\lambda_{\rm J}\sim\lambda_{\rm T}\sim H_{\rm s}$ \citep{1902RSPTA.199....1J,1964ApJ...139.1217T,1964ApJ...140.1056O,2015MNRAS.448.1007B}. Consequently, old remnants are expected to encounter several environmental effects at roughly the same evolutionary stage.

Our results suggest that environmental effects become important once remnants have expanded to scales of order $\lambda_{\rm J}$ and decelerated to velocities comparable to the ambient velocity dispersion, $v_{\rm s}\sim\sigma$. Depending on the local environment, this transition occurs after a few Myr for interactions with nearby overdensities and after a few tens of Myr for vertical stratification and galactic shear. Modeling all of these processes simultaneously would require a time-dependent description of self-gravitating galactic substructure subject to differential rotation, approaching the complexity of full numerical simulations \citepalias{2025A&A...702A..12R}.

Despite this complexity, several general trends emerge. Vertical stratification alone rarely leads to permanent breakout: only sufficiently powerful starbursts satisfy the conditions given by Eqs.~\ref{eq:break-out1} and \ref{eq:break-out2}\citep{2025A&A...701L...5R}, whereas weaker remnants eventually stall and fall back as galactic fountain flows. At the same time, galactic shear and large-scale substructure jointly deform remnants into elongated, non-spherical morphologies. Because molecular clouds are themselves stretched by differential rotation, evolved SBs are expected to align preferentially with the surrounding filamentary ISM, producing pitch angles comparable to those observed for galactic filaments \citep{2024ApJ...975...39X}.

Finally, we have considered only the gravitational potential of the gas and stationary (or co-rotating) environments. In realistic galaxies, stellar and dark matter gravity, together with turbulent and anisotropic gas flows, are expected to introduce additional departures from the evolution in idealized uniform media, particularly once remnant velocities become comparable to the characteristic ISM velocity dispersion.

\section{Limitations and future directions}\label{sec:discussion}

FRANZ provides an efficient analytical framework for modeling blastwave evolution in arbitrary environments while remaining sufficiently simple to permit analytical insight. This makes it useful not only for rapid parameter studies but also for developing physical intuition. Nevertheless, the model has several important limitations.

The one-zone approximation precludes phenomena that depend on the internal structure of the remnant, including the evolution of the reverse shock \citep{1999ApJS..120..299T}, the pressure-driven snowplow phase \citep{2016MNRAS.460.2962H}, mixing across the contact discontinuity \citep{2015ApJ...802...99K, 2025ApJ...990...49G}, and blastwave implosion \citep{2024ApJ...965..168R}. Likewise, shell fragmentation by fluid instabilities \citep{1994ApJ...428..186V} cannot be captured directly, although the magnitude of the surface-normal may provide a useful diagnostic of shear-induced fragmentation. Future work is needed to validate this idea.

The fully local formulation based on the sector approximation enables arbitrary blastwave geometries at low computational cost, but neglects couplings between different parts of the shock surface that are included in existing models \citep[e.g.][]{1969JFM....35...53L,1987A&A...186..287T,1995RvMP...67..661B,2020A&A...644A..72P,2024ApJ...960...81J}. While future work is needed to assess the validity of this simplification, the successes showcased in this paper and the substantial improvements in terms of computational cost are encouraging.

At present, FRANZ includes hydrodynamics, gravity, radiative cooling and a simplified treatment of cosmic rays \citep{2025A&A...701L...5R}. Its modular formulation and explicit representation of the local surface geometry make it straightforward to incorporate additional physics, such as magnetic fields, turbulence, multiphase gas, chemistry, dust, and a multi-zone description that can resolve the dynamics of the blastwave interior and reverse shock.

Finally, the predicted evolution of the shock radius, velocity and swept-up mass can readily be coupled to secondary models of cosmic-ray acceleration \citep{2025A&A...693A.145C}, dust processing \citep{2025arXiv251224677V}, chemistry and emission \citep{2022ApJ...938...23K}. FRANZ therefore provides a flexible framework for exploring blastwave evolution and connecting analytical models to observations.

\section{Conclusions}\label{sec:summary}

In this paper we introduced FRANZ, a modular, analytical framework for one-zone blastwave dynamics that follows the evolution of the shock surface in arbitrary environments, offering great flexibility with explosion models. Building on the thin-shell and sector approximations, the model extends classical analytical blastwave theory to complex galactic environments while remaining computationally inexpensive and easily extensible. Based on the model, we have published a code package written in Julia alongside this paper, which is publicly available from official Julia Registries and github. 

We demonstrated that FRANZ reproduces a number of well-established theoretical limits in uniform media before applying it to quantify the effects of vertical stratification, galactic rotation and large-scale density structure on the evolution of SNRs. Our analysis leads to the following main conclusions:
\begin{enumerate}
\item Once gravity and radiative cooling are included, vertical stratification alone does not guarantee disk breakout. Single SNe and weak SBs stall after approximately one free-fall time before subsequently falling back toward the disk, whereas only sufficiently powerful SBs ($\gg 40$ SNe within 1 Myr) are able to break out of the ISM. 

\item Galactic shear becomes dynamically important after a quarter epicycle, a timescale that is much longer than the typical timescales for blastwave evolution in a typical galactic disk. SNRs are  stretched into prolate shapes with pitch angles of $\sim15-45^\circ$, undergo epicyclic momentum oscillations, that periodically convert between radial expansion and tangential motion. Epicyclic motion reduces the efficiency of momentum coupling in continuously driven blastwaves (Fig.~\ref{fig:momentum_shear}), which might play a role in regulating feedback in galactic centers, where the shorter orbital timescales are comparable to feedback timescales.

\item Large-scale density structure strongly affects both the morphology and dynamics of SBs. Interactions with dense filaments produce highly anisotropic remnants and confound the observationally assigned ambient densities, affecting the interpretation of their expansion history. Ages assigned to SBs based on models in uniform media, may be systematically biased high.
\end{enumerate}

The modular formulation of FRANZ makes it straightforward to incorporate additional physics, such as magnetic fields or cosmic rays. The framework provides a useful tool for identifying the physical mechanisms responsible for the observed morphology and dynamics of evolved SNRs and SBs, without the need of costly numerical simulations and may aid in the interpretation of observations in complex galactic environments as well as the development of improved subgrid prescriptions for stellar feedback in large-scale simulations.

\begin{acknowledgements}
We thank the anonymous referee for their insightful comments and suggestions that helped to improve the quality of this work.
Computations were performed on the scicomp system at the European Southern Observatory in Garching.
This research was funded by the Deutsche Forschungsgemeinschaft (DFG, German Research Foundation) under Germany's Excellence Strategy – EXC 2094/2 – 390783311.
I thank the developers of the following software and
packages that were used in this work: \textsc{Julia} v1.10.0~\citep{Julia-2017},
\textsc{Matplotlib} v3.3.2~\citep{Hunter:2007},
HEALPix v4.2.4~\citep{2021ascl.soft09028T}, and \textsc{FRANZ} v.0.1.1~\citep{romano_2026_21180994}
\end{acknowledgements}

\bibliographystyle{aa}
\bibliography{bibliography}

@ARTICLE{2024ApJ...961...32K,
       author = {{Kobashi}, Ryosuke and {Lee}, Shiu-Hang and {Tanaka}, Takaaki and {Maeda}, Keiichi},
        title = "{Exploring the Circumstellar Environment of Tycho's Supernova Remnant. I. The Hydrodynamic Evolution of the Shock}",
      journal = {\apj},
         year = 2024,
        month = jan,
       volume = {961},
       number = {1},
          eid = {32},
        pages = {32},
          doi = {10.3847/1538-4357/ad05c2},
archivePrefix = {arXiv},
       eprint = {2310.14841},
 primaryClass = {astro-ph.HE},
       adsurl = {https://ui.adsabs.harvard.edu/abs/2024ApJ...961...32K}
}

@ARTICLE{2024ApJ...976L...4D,
       author = {{De Looze}, Ilse and {Milisavljevic}, Dan and {Temim}, Tea and {Dickinson}, Danielle and {Fesen}, Robert and {Arendt}, Richard G. and {Chastenet}, Jeremy and {Orlando}, Salvatore and {Vink}, Jacco and {Barlow}, Michael J. and {Kirchschlager}, Florian and {Priestley}, Felix D. and {Raymond}, John C. and {Rho}, Jeonghee and {Sartorio}, Nina S. and {Scheffler}, Tassilo and {Schmidt}, Franziska and {Blair}, William P. and {Fox}, Ori and {Fryer}, Christopher and {Janka}, Hans-Thomas and {Koo}, Bon-Chul and {Laming}, J. Martin and {Matsuura}, Mikako and {Patnaude}, Dan and {Rela{\~n}o}, M{\'o}nica and {Rest}, Armin and {Schmidt}, Judy and {Smith}, Nathan and {Sravan}, Niharika},
        title = "{The Green Monster Hiding in Front of Cas A: JWST Reveals a Dense and Dusty Circumstellar Structure Pockmarked by Ejecta Interactions}",
      journal = {\apjl},
         year = 2024,
        month = nov,
       volume = {976},
       number = {1},
          eid = {L4},
        pages = {L4},
          doi = {10.3847/2041-8213/ad855d},
archivePrefix = {arXiv},
       eprint = {2410.05402},
 primaryClass = {astro-ph.HE},
       adsurl = {https://ui.adsabs.harvard.edu/abs/2024ApJ...976L...4D}
}

@ARTICLE{2018A&A...612A.110A,
       author = {{Arias}, M. and {Vink}, J. and {de Gasperin}, F. and {Salas}, P. and {Oonk}, J.~B.~R. and {van Weeren}, R.~J. and {van Amesfoort}, A.~S. and {Anderson}, J. and {Beck}, R. and {Bell}, M.~E. and {Bentum}, M.~J. and {Best}, P. and {Blaauw}, R. and {Breitling}, F. and {Broderick}, J.~W. and {Brouw}, W.~N. and {Br{\"u}ggen}, M. and {Butcher}, H.~R. and {Ciardi}, B. and {de Geus}, E. and {Deller}, A. and {van Dijk}, P.~C.~G. and {Duscha}, S. and {Eisl{\"o}ffel}, J. and {Garrett}, M.~A. and {Grie{\ss}meier}, J.~M. and {Gunst}, A.~W. and {van Haarlem}, M.~P. and {Heald}, G. and {Hessels}, J. and {H{\"o}randel}, J. and {Holties}, H.~A. and {van der Horst}, A.~J. and {Iacobelli}, M. and {Juette}, E. and {Krankowski}, A. and {van Leeuwen}, J. and {Mann}, G. and {McKay-Bukowski}, D. and {McKean}, J.~P. and {Mulder}, H. and {Nelles}, A. and {Orru}, E. and {Paas}, H. and {Pandey-Pommier}, M. and {Pandey}, V.~N. and {Pekal}, R. and {Pizzo}, R. and {Polatidis}, A.~G. and {Reich}, W. and {R{\"o}ttgering}, H.~J.~A. and {Rothkaehl}, H. and {Schwarz}, D.~J. and {Smirnov}, O. and {Soida}, M. and {Steinmetz}, M. and {Tagger}, M. and {Thoudam}, S. and {Toribio}, M.~C. and {Vocks}, C. and {van der Wiel}, M.~H.~D. and {Wijers}, R.~A.~M.~J. and {Wucknitz}, O. and {Zarka}, P. and {Zucca}, P.},
        title = "{Low-frequency radio absorption in Cassiopeia A}",
      journal = {\aap},
         year = 2018,
        month = apr,
       volume = {612},
          eid = {A110},
        pages = {A110},
          doi = {10.1051/0004-6361/201732411},
archivePrefix = {arXiv},
       eprint = {1801.04887},
 primaryClass = {astro-ph.HE},
       adsurl = {https://ui.adsabs.harvard.edu/abs/2018A&A...612A.110A}
}

@ARTICLE{2023MNRAS.518.2320D,
       author = {{Deng}, Yunwei and {Zhang}, Zhi-Yu and {Zhou}, Ping and {Wang}, Junzhi and {Fang}, Min and {Lin}, Lingrui and {Bian}, Fuyan and {Chen}, Zhiwei and {Shi}, Yong and {Chen}, Guoyin and {Li}, Hui},
        title = "{Multiple gas phases in supernova remnant IC 443: mapping shocked H$_{2}$ with VLT/KMOS}",
      journal = {\mnras},
         year = 2023,
        month = jan,
       volume = {518},
       number = {2},
        pages = {2320-2340},
          doi = {10.1093/mnras/stac3139},
archivePrefix = {arXiv},
       eprint = {2210.16909},
 primaryClass = {astro-ph.GA},
       adsurl = {https://ui.adsabs.harvard.edu/abs/2023MNRAS.518.2320D}
}

@ARTICLE{2024MNRAS.52711685P,
       author = {{Payl{\i}}, G. and {Bak{\i}{\c{s}}}, H. and {Aktekin}, E. and {Sano}, H. and {Sezer}, A.},
        title = "{Discovery of optical emission from the supernova remnant G108.2-0.6 and its atomic environment}",
      journal = {\mnras},
         year = 2024,
        month = feb,
       volume = {527},
       number = {4},
        pages = {11685-11693},
          doi = {10.1093/mnras/stad3943},
archivePrefix = {arXiv},
       eprint = {2312.12862},
 primaryClass = {hep-ph},
       adsurl = {https://ui.adsabs.harvard.edu/abs/2024MNRAS.52711685P}
}

@ARTICLE{2023ApJ...944L..24W,
       author = {{Watkins}, Elizabeth J. and {Barnes}, Ashley T. and {Henny}, Kiana and {Kim}, Hwihyun and {Kreckel}, Kathryn and {Meidt}, Sharon E. and {Klessen}, Ralf S. and {Glover}, Simon C.~O. and {Williams}, Thomas G. and {Keller}, Benjamin W. and {Leroy}, Adam K. and {Rosolowsky}, Erik and {Lee}, Janice C. and {Anand}, Gagandeep S. and {Belfiore}, Francesco and {Bigiel}, Frank and {Blanc}, Guillermo A. and {Boquien}, M{\'e}d{\'e}ric and {Cao}, Yixian and {Chandar}, Rupali and {Chen}, Ness Mayker and {Chevance}, M{\'e}lanie and {Congiu}, Enrico and {Dale}, Daniel A. and {Deger}, Sinan and {Egorov}, Oleg V. and {Emsellem}, Eric and {Faesi}, Christopher M. and {Grasha}, Kathryn and {Groves}, Brent and {Hassani}, Hamid and {Henshaw}, Jonathan D. and {Herrera}, Cinthya and {Hughes}, Annie and {Jeffreson}, Sarah and {Jim{\'e}nez-Donaire}, Mar{\'\i}a J. and {Koch}, Eric W. and {Kruijssen}, J.~M. Diederik and {Larson}, Kirsten L. and {Liu}, Daizhong and {Lopez}, Laura A. and {Pessa}, Ismael and {Pety}, J{\'e}r{\^o}me and {Querejeta}, Miguel and {Saito}, Toshiki and {Sandstrom}, Karin and {Scheuermann}, Fabian and {Schinnerer}, Eva and {Sormani}, Mattia C. and {Stuber}, Sophia K. and {Thilker}, David A. and {Usero}, Antonio and {Whitmore}, Bradley C.},
        title = "{PHANGS-JWST First Results: A Statistical View on Bubble Evolution in NGC 628}",
      journal = {\apjl},
         year = 2023,
        month = feb,
       volume = {944},
       number = {2},
          eid = {L24},
        pages = {L24},
          doi = {10.3847/2041-8213/aca6e4},
archivePrefix = {arXiv},
       eprint = {2212.00811},
 primaryClass = {astro-ph.GA},
       adsurl = {https://ui.adsabs.harvard.edu/abs/2023ApJ...944L..24W}
}

@ARTICLE{2023ApJ...942...94Z,
       author = {{Zhang}, Shaobo and {Tian}, Wenwu and {Zhang}, Mengfei and {Zhu}, Hui and {Cui}, Xiaohong},
        title = "{Magnetohydrodynamic Simulations of the Supernova Remnant G1.9+0.3}",
      journal = {\apj},
         year = 2023,
        month = jan,
       volume = {942},
       number = {2},
          eid = {94},
        pages = {94},
          doi = {10.3847/1538-4357/aca7bf},
archivePrefix = {arXiv},
       eprint = {2211.12426},
 primaryClass = {astro-ph.HE},
       adsurl = {https://ui.adsabs.harvard.edu/abs/2023ApJ...942...94Z}
}

@ARTICLE{2023MNRAS.519.5358V,
       author = {{Vel{\'a}zquez}, P.~F. and {Meyer}, D.~M. -A. and {Chiotellis}, A. and {Cruz-{\'A}lvarez}, A.~E. and {Schneiter}, E.~M. and {Toledo-Roy}, J.~C. and {Reynoso}, E.~M. and {Esquivel}, A.},
        title = "{The sculpting of rectangular and jet-like morphologies in supernova remnants by anisotropic equatorially confined progenitor stellar winds}",
      journal = {\mnras},
         year = 2023,
        month = mar,
       volume = {519},
       number = {4},
        pages = {5358-5372},
          doi = {10.1093/mnras/stad039},
archivePrefix = {arXiv},
       eprint = {2301.03660},
 primaryClass = {astro-ph.HE},
       adsurl = {https://ui.adsabs.harvard.edu/abs/2023MNRAS.519.5358V}
}

@ARTICLE{2024arXiv240313641D,
       author = {{Duffell}, Paul C. and {Polin}, Abigail and {Mandal}, Soham},
        title = "{Sculpting the Morphology of Supernova Remnant Pa 30 via Efficient Ejecta Cooling}",
      journal = {arXiv e-prints},
         year = 2024,
        month = mar,
          eid = {arXiv:2403.13641},
        pages = {arXiv:2403.13641},
          doi = {10.48550/arXiv.2403.13641},
archivePrefix = {arXiv},
       eprint = {2403.13641},
 primaryClass = {astro-ph.HE},
       adsurl = {https://ui.adsabs.harvard.edu/abs/2024arXiv240313641D}
}

@ARTICLE{2006A&A...452L...1B,
       author = {{Breitschwerdt}, D. and {de Avillez}, M.~A.},
        title = "{The history and future of the Local and Loop I bubbles}",
      journal = {\aap},
         year = 2006,
        month = jun,
       volume = {452},
       number = {1},
        pages = {L1-L5},
          doi = {10.1051/0004-6361:20064989},
archivePrefix = {arXiv},
       eprint = {astro-ph/0604162},
 primaryClass = {astro-ph},
       adsurl = {https://ui.adsabs.harvard.edu/abs/2006A&A...452L...1B}
}

@ARTICLE{2016Natur.532...73B,
       author = {{Breitschwerdt}, D. and {Feige}, J. and {Schulreich}, M.~M. and {Avillez}, M.~A. De. and {Dettbarn}, C. and {Fuchs}, B.},
        title = "{The locations of recent supernovae near the Sun from modelling $^{60}$Fe transport}",
      journal = {\nat},
         year = 2016,
        month = apr,
       volume = {532},
       number = {7597},
        pages = {73-76},
          doi = {10.1038/nature17424},
       adsurl = {https://ui.adsabs.harvard.edu/abs/2016Natur.532...73B}
}

@ARTICLE{1988RvMP...60....1O,
       author = {{Ostriker}, Jeremiah P. and {McKee}, Christopher F.},
        title = "{Astrophysical blastwaves}",
      journal = {Reviews of Modern Physics},
         year = 1988,
        month = jan,
       volume = {60},
       number = {1},
        pages = {1-68},
          doi = {10.1103/RevModPhys.60.1},
       adsurl = {https://ui.adsabs.harvard.edu/abs/1988RvMP...60....1O}
}

@ARTICLE{1999ApJS..120..299T,
       author = {{Truelove}, J. Kelly and {McKee}, Christopher F.},
        title = "{Evolution of Nonradiative Supernova Remnants}",
      journal = {\apjs},
         year = 1999,
        month = feb,
       volume = {120},
       number = {2},
        pages = {299-326},
          doi = {10.1086/313176},
       adsurl = {https://ui.adsabs.harvard.edu/abs/1999ApJS..120..299T}
}

@ARTICLE{2019MNRAS.490.1961E,
       author = {{El-Badry}, Kareem and {Ostriker}, Eve C. and {Kim}, Chang-Goo and {Quataert}, Eliot and {Weisz}, Daniel R.},
        title = "{Evolution of supernovae-driven superbubbles with conduction and cooling}",
      journal = {\mnras},
         year = 2019,
        month = dec,
       volume = {490},
       number = {2},
        pages = {1961-1990},
          doi = {10.1093/mnras/stz2773},
archivePrefix = {arXiv},
       eprint = {1902.09547},
 primaryClass = {astro-ph.GA},
       adsurl = {https://ui.adsabs.harvard.edu/abs/2019MNRAS.490.1961E}
}

@ARTICLE{1969JFM....35...53L,
       author = {{Laumbach}, D.~D. and {Probstein}, R.~F.},
        title = "{A point explosion in a cold exponential atmosphere}",
      journal = {Journal of Fluid Mechanics},
         year = 1969,
        month = jan,
       volume = {35},
        pages = {53-75},
          doi = {10.1017/S0022112069000966},
       adsurl = {https://ui.adsabs.harvard.edu/abs/1969JFM....35...53L}
}

@ARTICLE{2016MNRAS.460.2962H,
       author = {{Haid}, S. and {Walch}, S. and {Naab}, T. and {Seifried}, D. and {Mackey}, J. and {Gatto}, A.},
        title = "{Supernova blast waves in wind-blown bubbles, turbulent, and power-law ambient media}",
      journal = {\mnras},
         year = 2016,
        month = aug,
       volume = {460},
       number = {3},
        pages = {2962-2978},
          doi = {10.1093/mnras/stw1082},
archivePrefix = {arXiv},
       eprint = {1604.04395},
 primaryClass = {astro-ph.GA},
       adsurl = {https://ui.adsabs.harvard.edu/abs/2016MNRAS.460.2962H}
}

@ARTICLE{2024ApJ...960...81J,
       author = {{Jim{\'e}nez}, S. and {Silich}, S. and {Mayya}, Y.~D. and {Zaragoza-Cardiel}, J.},
        title = "{What Holes in the Gas Distribution of Nearly Face-on Galaxies Can Tell Us about the Host Disk Parameters: The Case of the NGC 628 Southeast Superbubble}",
      journal = {\apj},
         year = 2024,
        month = jan,
       volume = {960},
       number = {1},
          eid = {81},
        pages = {81},
          doi = {10.3847/1538-4357/ad0cb8},
archivePrefix = {arXiv},
       eprint = {2311.08178},
 primaryClass = {astro-ph.GA},
       adsurl = {https://ui.adsabs.harvard.edu/abs/2024ApJ...960...81J}
}

@ARTICLE{2025MNRAS.540.1124L,
       author = {{Lau}, Cheryl S.~C. and {Bonnell}, Ian A.},
        title = "{Semi-confined supernova feedback in H II region bubbles}",
      journal = {\mnras},
         year = 2025,
        month = jun,
       volume = {540},
       number = {1},
        pages = {1124-1143},
          doi = {10.1093/mnras/staf756},
archivePrefix = {arXiv},
       eprint = {2410.21255},
 primaryClass = {astro-ph.GA},
       adsurl = {https://ui.adsabs.harvard.edu/abs/2025MNRAS.540.1124L}
}

@ARTICLE{2015ApJ...802...99K,
       author = {{Kim}, Chang-Goo and {Ostriker}, Eve C.},
        title = "{Momentum Injection by Supernovae in the Interstellar Medium}",
      journal = {\apj},
         year = 2015,
        month = apr,
       volume = {802},
       number = {2},
          eid = {99},
        pages = {99},
          doi = {10.1088/0004-637X/802/2/99},
archivePrefix = {arXiv},
       eprint = {1410.1537},
 primaryClass = {astro-ph.GA},
       adsurl = {https://ui.adsabs.harvard.edu/abs/2015ApJ...802...99K}
}

@ARTICLE{2022ApJS..262....9O,
       author = {{Oku}, Yuri and {Tomida}, Kengo and {Nagamine}, Kentaro and {Shimizu}, Ikkoh and {Cen}, Renyue},
        title = "{Osaka Feedback Model. II. Modeling Supernova Feedback Based on High-resolution Simulations}",
      journal = {\apjs},
         year = 2022,
        month = sep,
       volume = {262},
       number = {1},
          eid = {9},
        pages = {9},
          doi = {10.3847/1538-4365/ac77ff},
archivePrefix = {arXiv},
       eprint = {2201.00970},
 primaryClass = {astro-ph.GA},
       adsurl = {https://ui.adsabs.harvard.edu/abs/2022ApJS..262....9O}
}

@ARTICLE{1950RSPSA.201..159T,
       author = {{Taylor}, Geoffrey},
        title = "{The Formation of a Blast Wave by a Very Intense Explosion. I. Theoretical Discussion}",
      journal = {Proceedings of the Royal Society of London Series A},
         year = 1950,
        month = mar,
       volume = {201},
       number = {1065},
        pages = {159-174},
          doi = {10.1098/rspa.1950.0049},
       adsurl = {https://ui.adsabs.harvard.edu/abs/1950RSPSA.201..159T}
}

@BOOK{1959sdmm.book.....S,
       author = {{Sedov}, L.~I.},
        title = "{Similarity and Dimensional Methods in Mechanics}",
         year = 1959,
       adsurl = {https://ui.adsabs.harvard.edu/abs/1959sdmm.book.....S},
      publisher = {New York: Academic}
}

@ARTICLE{1988ApJ...334..252C,
       author = {{Cioffi}, Denis F. and {McKee}, Christopher F. and {Bertschinger}, Edmund},
        title = "{Dynamics of Radiative Supernova Remnants}",
      journal = {\apj},
         year = 1988,
        month = nov,
       volume = {334},
        pages = {252},
          doi = {10.1086/166834},
       adsurl = {https://ui.adsabs.harvard.edu/abs/1988ApJ...334..252C}
}

@ARTICLE{2016MNRAS.456..710F,
       author = {{Fierlinger}, Katharina M. and {Burkert}, Andreas and {Ntormousi}, Evangelia and {Fierlinger}, Peter and {Schartmann}, Marc and {Ballone}, Alessandro and {Krause}, Martin G.~H. and {Diehl}, Roland},
        title = "{Stellar feedback efficiencies: supernovae versus stellar winds}",
      journal = {\mnras},
         year = 2016,
        month = feb,
       volume = {456},
       number = {1},
        pages = {710-730},
          doi = {10.1093/mnras/stv2699},
archivePrefix = {arXiv},
       eprint = {1511.05151},
 primaryClass = {astro-ph.GA},
       adsurl = {https://ui.adsabs.harvard.edu/abs/2016MNRAS.456..710F}
}

@ARTICLE{1992ApJ...392..131S,
       author = {{Slavin}, Jonathan D. and {Cox}, Donald P.},
        title = "{Completing the Evolution of Supernova Remnants and Their Bubbles}",
      journal = {\apj},
         year = 1992,
        month = jun,
       volume = {392},
        pages = {131},
          doi = {10.1086/171412},
       adsurl = {https://ui.adsabs.harvard.edu/abs/1992ApJ...392..131S}
}

@ARTICLE{2024ApJ...965..168R,
       author = {{Romano}, Leonard E.~C. and {Behrendt}, Manuel and {Burkert}, Andreas},
        title = "{Cloud Formation by Supernova Implosion}",
      journal = {\apj},
         year = 2024,
        month = apr,
       volume = {965},
       number = {2},
          eid = {168},
        pages = {168},
          doi = {10.3847/1538-4357/ad2c05},
archivePrefix = {arXiv},
       eprint = {2402.05796},
 primaryClass = {astro-ph.GA},
       adsurl = {https://ui.adsabs.harvard.edu/abs/2024ApJ...965..168R}
}

@ARTICLE{1960SPhD....5...46K,
       author = {{Kompaneets}, A.~S.},
        title = "{A Point Explosion in an Inhomogeneous Atmosphere}",
      journal = {Soviet Physics Doklady},
         year = 1960,
        month = jul,
       volume = {5},
        pages = {46},
       adsurl = {https://ui.adsabs.harvard.edu/abs/1960SPhD....5...46K}
}

@ARTICLE{1976A&A....50..105M,
       author = {{Moellenhoff}, C.},
        title = "{An explosion model for extragalactic double radio sources.}",
      journal = {\aap},
         year = 1976,
        month = jul,
       volume = {50},
       number = {1},
        pages = {105-112},
       adsurl = {https://ui.adsabs.harvard.edu/abs/1976A&A....50..105M}
}

@ARTICLE{1990ApJ...354..513K,
       author = {{Koo}, Bon-Chul and {McKee}, Christopher F.},
        title = "{Dynamics of Adiabatic Blast Waves in Media of Finite Mass}",
      journal = {\apj},
         year = 1990,
        month = may,
       volume = {354},
        pages = {513},
          doi = {10.1086/168712},
       adsurl = {https://ui.adsabs.harvard.edu/abs/1990ApJ...354..513K}
}

@ARTICLE{1987A&A...186..287T,
       author = {{Tenorio-Tagle}, G. and {Palous}, J.},
        title = "{Giant-scale supernova remnants - The role of differential galactic rotation and the formation of molecular clouds}",
      journal = {\aap},
         year = 1987,
        month = nov,
       volume = {186},
       number = {1-2},
        pages = {287-294},
       adsurl = {https://ui.adsabs.harvard.edu/abs/1987A&A...186..287T}
}

@ARTICLE{1995RvMP...67..661B,
       author = {{Bisnovatyi-Kogan}, G.~S. and {Silich}, S.~A.},
        title = "{Shock-wave propagation in the nonuniform interstellar medium}",
      journal = {Reviews of Modern Physics},
         year = 1995,
        month = jul,
       volume = {67},
       number = {3},
        pages = {661-712},
          doi = {10.1103/RevModPhys.67.661},
       adsurl = {https://ui.adsabs.harvard.edu/abs/1995RvMP...67..661B}
}

@ARTICLE{2020A&A...644A..72P,
       author = {{Palou{\v{s}}}, J. and {Ehlerov{\'a}}, S. and {W{\"u}nsch}, R. and {Morris}, M.~R.},
        title = "{Can supernova shells feed supermassive black holes in galactic nuclei?}",
      journal = {\aap},
         year = 2020,
        month = dec,
       volume = {644},
          eid = {A72},
        pages = {A72},
          doi = {10.1051/0004-6361/202038768},
archivePrefix = {arXiv},
       eprint = {2010.15412},
 primaryClass = {astro-ph.GA},
       adsurl = {https://ui.adsabs.harvard.edu/abs/2020A&A...644A..72P}
}

@ARTICLE{2025A&A...702A..12R,
       author = {{Romano}, Leonard E.~C. and {Behrendt}, Manuel and {Burkert}, Andreas},
        title = "{SISSI: Supernovae in a stratified, shearing interstellar medium: I. The geometry of supernova remnants}",
      journal = {\aap},
         year = 2025,
        month = oct,
       volume = {702},
          eid = {A12},
        pages = {A12},
          doi = {10.1051/0004-6361/202554571},
archivePrefix = {arXiv},
       eprint = {2503.12977},
 primaryClass = {astro-ph.GA},
       adsurl = {https://ui.adsabs.harvard.edu/abs/2025A&A...702A..12R}
}

@ARTICLE{1977ApJ...218..377W,
       author = {{Weaver}, R. and {McCray}, R. and {Castor}, J. and {Shapiro}, P. and {Moore}, R.},
        title = "{Interstellar bubbles. II. Structure and evolution.}",
      journal = {\apj},
         year = 1977,
        month = dec,
       volume = {218},
        pages = {377-395},
          doi = {10.1086/155692},
       adsurl = {https://ui.adsabs.harvard.edu/abs/1977ApJ...218..377W}
}

@ARTICLE{2024ApJ...970...18L,
       author = {{Lancaster}, Lachlan and {Ostriker}, Eve C. and {Kim}, Chang-Goo and {Kim}, Jeong-Gyu and {Bryan}, Greg L.},
        title = "{Geometry, Dissipation, Cooling, and the Dynamical Evolution of Wind-blown Bubbles}",
      journal = {\apj},
         year = 2024,
        month = jul,
       volume = {970},
       number = {1},
          eid = {18},
        pages = {18},
          doi = {10.3847/1538-4357/ad47f6},
archivePrefix = {arXiv},
       eprint = {2405.02396},
 primaryClass = {astro-ph.GA},
       adsurl = {https://ui.adsabs.harvard.edu/abs/2024ApJ...970...18L}
}

@ARTICLE{2005ApJ...622..759G,
       author = {{G{\'o}rski}, K.~M. and {Hivon}, E. and {Banday}, A.~J. and {Wandelt}, B.~D. and {Hansen}, F.~K. and {Reinecke}, M. and {Bartelmann}, M.},
        title = "{HEALPix: A Framework for High-Resolution Discretization and Fast Analysis of Data Distributed on the Sphere}",
      journal = {\apj},
         year = 2005,
        month = apr,
       volume = {622},
       number = {2},
        pages = {759-771},
          doi = {10.1086/427976},
archivePrefix = {arXiv},
       eprint = {astro-ph/0409513},
 primaryClass = {astro-ph},
       adsurl = {https://ui.adsabs.harvard.edu/abs/2005ApJ...622..759G}
}

@ARTICLE{2024ApJ...975...39X,
       author = {{Xie}, Yi-Heng and {Li}, Guang-Xing and {Chen}, Bing-Qiu},
        title = "{Morphology of Molecular Clouds at Kiloparsec Scale in the Milky Way: Shear-induced Alignment and Vertical Confinement}",
      journal = {\apj},
         year = 2024,
        month = nov,
       volume = {975},
       number = {1},
          eid = {39},
        pages = {39},
          doi = {10.3847/1538-4357/ad7378},
archivePrefix = {arXiv},
       eprint = {2408.14095},
 primaryClass = {astro-ph.GA},
       adsurl = {https://ui.adsabs.harvard.edu/abs/2024ApJ...975...39X}
}

@ARTICLE{2025A&A...701L...5R,
       author = {{Romano}, Leonard E.~C. and {Owen}, Ellis R. and {Nagamine}, Kentaro},
        title = "{Starburst-driven galactic outflows: Unveiling the suppressive role of cosmic ray halos}",
      journal = {\aap},
         year = 2025,
        month = sep,
       volume = {701},
          eid = {L5},
        pages = {L5},
          doi = {10.1051/0004-6361/202554590},
archivePrefix = {arXiv},
       eprint = {2503.13261},
 primaryClass = {astro-ph.GA},
       adsurl = {https://ui.adsabs.harvard.edu/abs/2025A&A...701L...5R}
}

@ARTICLE{2022ApJ...932...88O,
       author = {{Orr}, Matthew E. and {Fielding}, Drummond B. and {Hayward}, Christopher C. and {Burkhart}, Blakesley},
        title = "{Bursting Bubbles: Feedback from Clustered Supernovae and the Trade-off Between Turbulence and Outflows}",
      journal = {\apj},
         year = 2022,
        month = jun,
       volume = {932},
       number = {2},
          eid = {88},
        pages = {88},
          doi = {10.3847/1538-4357/ac6c26},
archivePrefix = {arXiv},
       eprint = {2109.14656},
 primaryClass = {astro-ph.GA},
       adsurl = {https://ui.adsabs.harvard.edu/abs/2022ApJ...932...88O}
}

@ARTICLE{2009ApJ...694L..16H,
       author = {{Hewitt}, John W. and {Yusef-Zadeh}, Farhad},
        title = "{Discovery of New Interacting Supernova Remnants in the Inner Galaxy}",
      journal = {\apjl},
         year = 2009,
        month = mar,
       volume = {694},
       number = {1},
        pages = {L16-L20},
          doi = {10.1088/0004-637X/694/1/L16},
archivePrefix = {arXiv},
       eprint = {0902.1386},
 primaryClass = {astro-ph.GA},
       adsurl = {https://ui.adsabs.harvard.edu/abs/2009ApJ...694L..16H}
}

@ARTICLE{2023ApJ...944..110M,
       author = {{Mayker Chen}, Ness and {Leroy}, Adam K. and {Lopez}, Laura A. and {Benincasa}, Samantha and {Chevance}, M{\'e}lanie and {Glover}, Simon C.~O. and {Hughes}, Annie and {Kreckel}, Kathryn and {Sarbadhicary}, Sumit and {Sun}, Jiayi and {Thompson}, Todd A. and {Utomo}, Dyas and {Bigiel}, Frank and {Blanc}, Guillermo A. and {Dale}, Daniel A. and {Grasha}, Kathryn and {Kruijssen}, J.~M. Diederik and {Pan}, Hsi-An and {Querejeta}, Miguel and {Schinnerer}, Eva and {Watkins}, Elizabeth J. and {Williams}, Thomas G.},
        title = "{Comparing the Locations of Supernovae to CO (2-1) Emission in Their Host Galaxies}",
      journal = {\apj},
         year = 2023,
        month = feb,
       volume = {944},
       number = {1},
          eid = {110},
        pages = {110},
          doi = {10.3847/1538-4357/acab00},
archivePrefix = {arXiv},
       eprint = {2212.09766},
 primaryClass = {astro-ph.GA},
       adsurl = {https://ui.adsabs.harvard.edu/abs/2023ApJ...944..110M}
}

@ARTICLE{2023MNRAS.523.1421M,
       author = {{Makarenko}, Ekaterina I. and {Walch}, Stefanie and {Clarke}, Seamus D. and {Seifried}, Daniel and {Naab}, Thorsten and {N{\"u}rnberger}, Pierre C. and {Rathjen}, Tim-Eric},
        title = "{How do supernova remnants cool? - I. Morphology, optical emission lines, and shocks}",
      journal = {\mnras},
         year = 2023,
        month = jul,
       volume = {523},
       number = {1},
        pages = {1421-1440},
          doi = {10.1093/mnras/stad1472},
archivePrefix = {arXiv},
       eprint = {2305.07652},
 primaryClass = {astro-ph.HE},
       adsurl = {https://ui.adsabs.harvard.edu/abs/2023MNRAS.523.1421M}
}

@ARTICLE{1902RSPTA.199....1J,
       author = {{Jeans}, J.~H.},
        title = "{The Stability of a Spherical Nebula}",
      journal = {Philosophical Transactions of the Royal Society of London Series A},
         year = 1902,
        month = jan,
       volume = {199},
        pages = {1-53},
          doi = {10.1098/rsta.1902.0012},
       adsurl = {https://ui.adsabs.harvard.edu/abs/1902RSPTA.199....1J}
}

@ARTICLE{1964ApJ...139.1217T,
       author = {{Toomre}, A.},
        title = "{On the gravitational stability of a disk of stars.}",
      journal = {\apj},
         year = 1964,
        month = may,
       volume = {139},
        pages = {1217-1238},
          doi = {10.1086/147861},
       adsurl = {https://ui.adsabs.harvard.edu/abs/1964ApJ...139.1217T}
}

@ARTICLE{1964ApJ...140.1056O,
       author = {{Ostriker}, J.},
        title = "{The Equilibrium of Polytropic and Isothermal Cylinders.}",
      journal = {\apj},
         year = 1964,
        month = oct,
       volume = {140},
        pages = {1056},
          doi = {10.1086/148005},
       adsurl = {https://ui.adsabs.harvard.edu/abs/1964ApJ...140.1056O}
}

@ARTICLE{2022Natur.601..334Z,
       author = {{Zucker}, Catherine and {Goodman}, Alyssa A. and {Alves}, Jo{\~a}o and {Bialy}, Shmuel and {Foley}, Michael and {Speagle}, Joshua S. and {Gro{\^I}{\texttwosuperior}schedl}, Josefa and {Finkbeiner}, Douglas P. and {Burkert}, Andreas and {Khimey}, Diana and {Swiggum}, Cameren},
        title = "{Star formation near the Sun is driven by expansion of the Local Bubble}",
      journal = {\nat},
         year = 2022,
        month = jan,
       volume = {601},
       number = {7893},
        pages = {334-337},
          doi = {10.1038/s41586-021-04286-5},
archivePrefix = {arXiv},
       eprint = {2201.05124},
 primaryClass = {astro-ph.GA},
       adsurl = {https://ui.adsabs.harvard.edu/abs/2022Natur.601..334Z}
}

@ARTICLE{2026A&A...707A..98R,
       author = {{Romano}, Leonard E.~C. and {Burkert}, Andreas},
        title = "{SISSI: Supernovae in a stratified, shearing interstellar medium: II. Star formation near the Sun is quenched by expansion of the Local Bubble}",
      journal = {\aap},
         year = 2026,
        month = mar,
       volume = {707},
          eid = {A98},
        pages = {A98},
          doi = {10.1051/0004-6361/202558080},
archivePrefix = {arXiv},
       eprint = {2509.04221},
 primaryClass = {astro-ph.GA},
       adsurl = {https://ui.adsabs.harvard.edu/abs/2026A&A...707A..98R}
}

@ARTICLE{2015MNRAS.448.1007B,
       author = {{Behrendt}, M. and {Burkert}, A. and {Schartmann}, M.},
        title = "{Structure formation in gas-rich galactic discs with finite thickness: from discs to rings}",
      journal = {\mnras},
         year = 2015,
        month = mar,
       volume = {448},
       number = {1},
        pages = {1007-1019},
          doi = {10.1093/mnras/stv027},
archivePrefix = {arXiv},
       eprint = {1408.5902},
 primaryClass = {astro-ph.GA},
       adsurl = {https://ui.adsabs.harvard.edu/abs/2015MNRAS.448.1007B}
}

@ARTICLE{1994ApJ...428..186V,
       author = {{Vishniac}, Ethan T.},
        title = "{Nonlinear instabilities in shock-bounded slabs}",
      journal = {\apj},
         year = 1994,
        month = jun,
       volume = {428},
       number = {1},
        pages = {186-208},
          doi = {10.1086/174231},
archivePrefix = {arXiv},
       eprint = {astro-ph/9306025},
 primaryClass = {astro-ph},
       adsurl = {https://ui.adsabs.harvard.edu/abs/1994ApJ...428..186V}
}

@ARTICLE{2025A&A...693A.145C,
       author = {{Cristofari}, P. and {Tatischeff}, V. and {Chabot}, M.},
        title = "{Diffusive shock acceleration of dust grains at supernova remnants}",
      journal = {\aap},
         year = 2025,
        month = jan,
       volume = {693},
          eid = {A145},
        pages = {A145},
          doi = {10.1051/0004-6361/202452436},
archivePrefix = {arXiv},
       eprint = {2410.23190},
 primaryClass = {astro-ph.HE},
       adsurl = {https://ui.adsabs.harvard.edu/abs/2025A&A...693A.145C}
}

@ARTICLE{2025arXiv251224677V,
       author = {{Vasiliev}, Evgenii O.},
        title = "{Destruction of the interstellar dust by a supernova}",
      journal = {arXiv e-prints},
         year = 2025,
        month = dec,
          eid = {arXiv:2512.24677},
        pages = {arXiv:2512.24677},
          doi = {10.48550/arXiv.2512.24677},
archivePrefix = {arXiv},
       eprint = {2512.24677},
 primaryClass = {astro-ph.GA},
       adsurl = {https://ui.adsabs.harvard.edu/abs/2025arXiv251224677V}
}

@ARTICLE{2022ApJ...938...23K,
       author = {{Katsuragawa}, Miho and {Lee}, Shiu-Hang and {Odaka}, Hirokazu and {Bamba}, Aya and {Matsumura}, Hideaki and {Takahashi}, Tadayuki},
        title = "{On the Formation of Over-ionized Plasma in Evolved Supernova Remnants}",
      journal = {\apj},
         year = 2022,
        month = oct,
       volume = {938},
       number = {1},
          eid = {23},
        pages = {23},
          doi = {10.3847/1538-4357/ac8cf9},
archivePrefix = {arXiv},
       eprint = {2208.12451},
 primaryClass = {astro-ph.HE},
       adsurl = {https://ui.adsabs.harvard.edu/abs/2022ApJ...938...23K}
}

@ARTICLE{2025ApJ...990...49G,
       author = {{Guo}, Minghao and {Kim}, Chang-Goo and {Stone}, James M.},
        title = "{Evolution of Supernova Remnants in a Cloudy Multiphase Interstellar Medium}",
      journal = {\apj},
         year = 2025,
        month = sep,
       volume = {990},
       number = {1},
          eid = {49},
        pages = {49},
          doi = {10.3847/1538-4357/adeb85},
archivePrefix = {arXiv},
       eprint = {2411.12809},
 primaryClass = {astro-ph.GA},
       adsurl = {https://ui.adsabs.harvard.edu/abs/2025ApJ...990...49G}
}

@ARTICLE{2019MNRAS.483.4551E,
       author = {{Elwood}, Benjamin D. and {Murphy}, Jeremiah W. and {D{\'\i}az-Rodr{\'\i}guez}, Mariangelly},
        title = "{Inferring properties of the ISM from supernova remnant size distributions}",
      journal = {\mnras},
         year = 2019,
        month = mar,
       volume = {483},
       number = {4},
        pages = {4551-4559},
          doi = {10.1093/mnras/sty3355},
       adsurl = {https://ui.adsabs.harvard.edu/abs/2019MNRAS.483.4551E}
}

@article{Julia-2017,
    title={Julia: A fresh approach to numerical computing},
    author={Bezanson, Jeff and Edelman, Alan and Karpinski, Stefan and Shah, Viral B},
    journal={SIAM {R}eview},
    volume={59},
    number={1},
    pages={65--98},
    year={2017},
    publisher={SIAM},
    doi={10.1137/141000671},
    url={https://epubs.siam.org/doi/10.1137/141000671}
}

@Article{Hunter:2007,
  Author    = {Hunter, J. D.},
  Title     = {Matplotlib: A 2D graphics environment},
  Journal   = {Computing in Science \& Engineering},
  Volume    = {9},
  Number    = {3},
  Pages     = {90--95},
  publisher = {IEEE COMPUTER SOC},
  doi       = {10.1109/MCSE.2007.55},
  year      = 2007
}

@MISC{2021ascl.soft09028T,
       author = {{Tomasi}, Maurizio and {Li}, Zack},
        title = "{Healpix.jl: Julia-only port of the HEALPix library}",
      version = {2.3.0},     
         year = 2021,
        month = sep,
          eid = {ascl:2109.028},
        pages = {ascl:2109.028},
archivePrefix = {ascl},
       eprint = {2109.028},
       adsurl = {https://ui.adsabs.harvard.edu/abs/2021ascl.soft09028T}
}

@software{romano_2026_21180994,
  author       = {Romano, Leonard Elias Cornelius},
  title        = {leonardromano/FRANZ.jl: v0.1.0},
  month        = jul,
  year         = 2026,
  publisher    = {Zenodo},
  version      = {v0.1.0},
  doi          = {10.5281/zenodo.21180994},
  url          = {https://doi.org/10.5281/zenodo.21180994},
}

@ARTICLE{1974ApJ...188..501C,
       author = {{Chevalier}, Roger A.},
        title = "{The Evolution of Supernova Remnants. Spherically Symmetric Models}",
      journal = {\apj},
         year = 1974,
        month = mar,
       volume = {188},
        pages = {501-516},
          doi = {10.1086/152740},
       adsurl = {https://ui.adsabs.harvard.edu/abs/1974ApJ...188..501C}
}

@ARTICLE{1981MNRAS.195.1011F,
       author = {{Falle}, S.~A.~E.~G.},
        title = "{Catastrophic cooling in supernova remnants}",
      journal = {\mnras},
         year = 1981,
        month = jun,
       volume = {195},
        pages = {1011-1028},
          doi = {10.1093/mnras/195.4.1011},
       adsurl = {https://ui.adsabs.harvard.edu/abs/1981MNRAS.195.1011F}
}

\begin{appendix}

\section{Derivation of the pressure-gradient force}\label{app:pressure-gradient}

We derive the expression for the pressure-gradient force Eq. \ref{eq:pressure_gradient} following the formalism of \citet{1969JFM....35...53L}.
The expression for the pressure gradient can be read off from their final result, by noting that they consider a strong adiabatic blastwave, without central forcing, that expands into a stratified atmosphere along nearly straight streamlines, neglecting the effects of gravity.
Under these assumptions the equation of motion Eq. \ref{eq:EoM} simplifies to
\begin{equation}
    \frac{\text{d}}{\text{d}t}\left(M v_{s}\right) = \Delta P r_{\text{s}}^2 ~.
\end{equation}

The starting point of the derivation is the Taylor-expansion of the Eulerian radial coordinate $r$ in terms of the Lagrangian radial coordinate $r_{0}$ at $t=0$, near the radial location of the shock front $r_{s}$ to second order
\begin{equation}\label{eq:TaylorExpansion}
    r = r_{\text{s}} + \left.\frac{\partial r}{\partial r_0}\right|_{r_{\text{s}}} \left(r_0 - r_{\text{s}}\right) + \frac{1}{2} \left.\frac{\partial^2 r}{\partial r_0^2}\right|_{r_{\text{s}}} \left(r_0 - r_{\text{s}}\right)^2 +... ~.
\end{equation}
In order to derive the equation of motion, we make use of the thin-shell approximation, in which $r \rightarrow r_{\text{s}}$ is independent of $r_{0}$. 
In this approximation we can approximate the integral of the Euler-equation for the momentum evolution
\begin{equation}
    \frac{\partial^2 r}{\partial t^2} = - \frac{r^2}{\rho_{0} r_0^2} \frac{\partial P}{\partial r_0} \label{eq:momentum_equation}~,
\end{equation}
using the value of the acceleration at the shock
\begin{eqnarray}
    P\left(0, t\right) - P_{\text{s}}\left(r_{\text{s}}\right) &=& \int_{0}^{r_\text{{s}}} \frac{1}{r^2} \frac{\partial^2 r}{\partial t^2} \rho_{0} r_{0}^2 \text{d}r_{0} \nonumber\\ &\approx& \frac{1}{r_{\text{s}}^2} \left.\frac{\partial^2 r}{\partial t^2}\right|_{r_{\text{s}}} M\left(r_{\text{s}}\right) ~, \label{eq:momentum_integrated}
\end{eqnarray}
and similarly we can approximate the energy integral using the value of the kinetic energy density at the shock
\begin{eqnarray}
    E\left(r_{\text{s}}\right) &=& \int_{0}^{r_{\text{s}}} \frac{P\left(r\right)}{\gamma-1} r^2 \text{d}r + \int_0^{r_{\text{s}}} \frac{1}{2} \left(\frac{\partial r}{\partial t}\right)^2 \rho_0 r_0^2 \text{d}r_0 \nonumber\\
    &\approx& \frac{P\left(0\right) \, r_{\text{s}}^3}{3 \left(\gamma-1\right)} + \frac{1}{2}\, M\left(r_{\text{s}}\right) \left.\left(\frac{\partial r}{\partial t}\right)^2\right|_{r_{\text{s}}} ~. \label{eq:energy_integrated}
\end{eqnarray}
By combining Eqs. \ref{eq:momentum_integrated} and \ref{eq:energy_integrated} we can eliminate $P\left(0, t\right)$, which leaves us with the problem of deriving expressions for the various remaining quantities at the shock $r_{\text{s}}$.

\subsection{Derivation of the acceleration at the shock}

To leading order(s) the first two time-derivatives of Eq.~\ref{eq:TaylorExpansion} are
\begin{eqnarray}
    \left(\frac{\partial r}{\partial t}\right) &=& \left(1 - \left.\frac{\partial r}{\partial r_0}\right|_{r_{\text{s}}}\right) \, \dot{r}_{\text{s}} + \nonumber \\ &+& \left[\frac{\partial}{\partial t} \left(\left.\frac{\partial r}{\partial r_0}\right|_{r_{\text{s}}}\right) - \left.\frac{\partial^2 r}{\partial r_0^2}\right|_{r_{\text{s}}} \dot{r}_{\text{s}} \right] \,  \left(r_0 - r_{\text{s}}\right) + ... ~,\label{eq:velocity} \\
    \left(\frac{\partial^2 r}{\partial t^2}\right) &=& \left(1 - \left.\frac{\partial r}{\partial r_0}\right|_{r_{\text{s}}}\right) \, \ddot{r}_{\text{s}} - \nonumber\\ &-& 2 \, \frac{\partial}{\partial t} \left(\left.\frac{\partial r}{\partial r_0}\right|_{r_{\text{s}}}\right) \dot{r}_{\text{s}} + \left.\frac{\partial^2 r}{\partial r_0^2}\right|_{r_{\text{s}}} \dot{r}_{\text{s}}^2+ ... ~,\label{eq:acceleration}
\end{eqnarray}
where $\dot{r}_{\text{s}} = v_{\text{s}}$ is the shock speed.

The Eulerian and Lagrangian radial coordinates are related by the continuity equation
\begin{equation}
    \rho_{0} \, r_0^2 \text{d}r_0 = \rho r^2 \text{d}r ~,  \label{eq:continuity}  
\end{equation}
where $\rho_{0}\left(r_{0}\right)$ is the unperturbed density profile at $r_{0}$ and $\rho\left(r, t\right)$ is the density at $r$ and time $t$.
We can use this relation in combination with the Rankine-Hugoniot jump-conditions for strong shocks \citep[e.g.][]{1988RvMP...60....1O} to evaluate
\begin{equation}
    \left.\frac{\partial r}{\partial r_0}\right|_{r_{\text{s}}} = \frac{\gamma-1}{\gamma+1} = \chi^{-1},
\end{equation}
which is constant in time and determines the velocity at the shock
\begin{equation}
    \left.\frac{\partial r}{\partial t}\right|_{r_{\text{s}}} = \frac{2}{\gamma+1} v_{\text{s}} ~. \label{eq:speed_at_shock}
\end{equation}

By making use of the momentum equation \ref{eq:momentum_equation} and the Rankine-Hugoniot jump-condition for the pressure at the shock
\begin{equation}
    P_{\text{s}} = \frac{2}{\gamma+1} \rho_{0}\left(r_{\text{s}}\right)\, \dot{r}_{\text{s}}^2\, \label{eq:shock_pressure}
\end{equation}
we can evaluate Eq. \ref{eq:acceleration} at the shock to obtain
\begin{equation}
    - \left.\frac{\partial^2 r}{\partial r_0^2}\right|_{r_{\text{s}}} \dot{r}_{\text{s}}^2 = \frac{2}{\gamma+1} \dot{r}_{\text{s}}^2 \left.\left(\frac{1}{P} \frac{\partial P}{\partial r_{0}}\right)\right|_{r_{\text{s}}} + \frac{2}{\gamma+1} \ddot{r}_{\text{s}} ~. \label{eq:second_derivative}
\end{equation}

The pressure-gradient term can be further evaluated by utilizing the adiabatic condition
\begin{equation}
    \frac{P\left(r_{0}, t\right)}{P_{\text{s}}\left(r_{0}\right)} = \left(\frac{\rho\left(r_{0}, t\right)}{\rho_{\text{s}}\left(r_{0}\right)}\right)^{\gamma} ~, \label{eq:adiabatic}
\end{equation}
which yields for the pressure gradient at the shock
\begin{equation}
    \left.\left(\frac{1}{P} \frac{\partial P}{\partial r_{0}}\right)\right|_{r_{\text{s}}} = \left.\left(\frac{1}{P_{\text{s}}} \frac{\partial P_\text{s}}{\partial r_{0}} + \frac{\gamma}{\rho} \frac{\partial\rho}{\partial r_{0}} - \frac{\gamma}{\rho_{0}} \frac{\partial\rho_{0}}{\partial \rho_{0}}\right)\right|_{r_{\text{s}}}.
\end{equation}

The first term can be evaluated from Eq. \ref{eq:shock_pressure} yielding
\begin{equation}
    \left.\left(\frac{1}{P_{\text{s}}} \frac{\partial P_{\text{s}}}{\partial r_{0}}\right)\right|_{r_{\text{s}}} = \left.\left(\frac{1}{\rho_{0}} \frac{\partial \rho_{0}}{\partial r_{0}}\right)\right|_{r_{\text{s}}} + \frac{2 \ddot{r}_{\text{s}}}{\dot{r}_{\text{s}}^2} ~,
\end{equation}
since by the chain-rule
\begin{equation}
    \left. \left(\frac{\partial \dot{r}_{\text{s}}}{\partial r_{0}}\right)\right|_{r_{\text{s}}} = \ddot{r}_{\text{s}} \left. \left(\frac{\partial t}{\partial r_{0}}\right)\right|_{r_{\text{s}}} = \ddot{r}_{\text{s}} \, \dot{r}_{\text{s}}^{-1} ~.
\end{equation}

The second term can be evaluated using the continuity equation \ref{eq:continuity} yielding
\begin{equation}
    \left.\left(\frac{1}{\rho} \frac{\partial\rho}{\partial r_{0}}\right)\right|_{r_{\text{s}}} =  \left.\frac{1}{\rho_{0}} \frac{\partial\rho_{0}}{\partial r_{0}}\right|_{r_{\text{s}}} + \frac{4}{\gamma+1} \, r_{\text{s}}^{-1} - \chi \left.\frac{\partial^2 r}{\partial r_0^2}\right|_{r_{\text{s}}} ~.
\end{equation}
We thus obtain for the pressure gradient
\begin{equation}
    \left.\left(\frac{1}{P} \frac{\partial P}{\partial r_{0}}\right)\right|_{r_{\text{s}}} = \left.\left(\frac{1}{\rho_{0}} \frac{\partial \rho_{0}}{\partial r_{0}}\right)\right|_{r_{\text{s}}} + \frac{2 \ddot{r}_{\text{s}}}{\dot{r}_{\text{s}}^2} + \frac{4\gamma}{\gamma+1}\,  r_{\text{s}}^{-1} - \gamma \, \chi \left.\frac{\partial^2 r}{\partial r_0^2}\right|_{r_{\text{s}}} ~,
\end{equation}
which we can plug into Eq. \ref{eq:second_derivative} and solve for $\left.\left(\partial^2 r / \partial r_0^2\right)\right|_{r_{\text{s}}}$ to obtain
\begin{equation}
    \left.\frac{\partial^2 r}{\partial r_0^2}\right|_{r_{\text{s}}} = - \frac{2\chi^{-1}}{\gamma+1} \frac{k_{\rho} - 3 \, r_{\text{s}}\, \ddot{r}_{\text{s}} / \dot{r}_{\text{s}}^2  - 4\gamma / \left(\gamma+1\right)}{r_{\text{s}}} ~.
\end{equation}
Finally, we can use this to obtain the acceleration at the shock
\begin{equation}
    \left.\left(\frac{\partial^2 r}{\partial^2 t}\right)\right|_{r_{\text{s}}} = \frac{4\left(2\gamma-1\right)}{\left(\gamma+1\right)^2} \ddot{r}_{\text{s}} - \frac{2\chi^{-1}}{\gamma+1} \left(k_{\rho} - \frac{4\gamma}{\left(\gamma+1\right)} \right) \, \frac{\dot{r}_{\text{s}}^2}{r_{\text{s}}} ~. \label{eq:acceleration_solved}
\end{equation}

\subsection{Obtaining the equation of motion}

We have now all the pieces in place to derive the equation of motion.
We combine Eqs. \ref{eq:momentum_integrated} and \ref{eq:energy_integrated} to eliminate $P\left(0, t\right)$, plug in Eqs. \ref{eq:shock_pressure}, \ref{eq:speed_at_shock} and \ref{eq:acceleration_solved} and solve for $M\dot{v}_{\text{s}}$ to obtain
\begin{eqnarray}
    M\dot{v}_{\text{s}} = \frac{\left(\gamma+1\right)^2}{4\left(2\gamma-1\right)} \left[3\left(\gamma-1\right) \frac{E}{r_{\text{s}}} - \frac{2}{\gamma+1} \rho_{0} r_{\text{s}}^2 v_{\text{s}}^2 + \right. \nonumber\\  \left. + 2 \frac{\gamma-1}{\left(\gamma+1\right)^2} \left(k_{\rho} - \left(3 + \frac{4\gamma}{\gamma+1}\right)\right) \, \frac{M \, v_{\text{s}}^2}{r_{\text{s}}}\right] ~,
\end{eqnarray}
adding $\dot{M}v_{\text{s}} = \rho_{0}\, r_{\text{s}}^2\, v_{\text{s}}^2$ finally yields the equation of motion
\begin{eqnarray}
    \frac{\text{d}}{\text{d} t}\left(M v_{\text{s}}\right) = \frac{\gamma-1}{2\left(2\gamma-1\right)} \, \left[ \frac{3\left(\gamma+1\right)^2}{2} \frac{E}{r_{\text{s}}^3} + 3 \rho_{0} v_{\text{s}}^2 \right.\nonumber\\ \left.+ \left(k_{\rho} - \left(3 + \frac{4\gamma}{\gamma+1}\right)\right) \, \frac{M \, v_{\text{s}}^2}{r_{\text{s}}^3}\right]\, r_{\text{s}}^2 ~,
\end{eqnarray}
from which the expression for the pressure gradient Eq. \ref{eq:pressure_gradient} can be read off.

While this derivation assumes an adiabatic blastwave in a stationary medium, it might be possible to lift this assumption and derive the pressure-gradient force in more complex cases, such as a radiatively cooling isothermal shock.
In this case the compression-ratio at the shock and the shock-pressure would need to be updated to those of an isothermal shock. 
Moreover, the adiabatic condition would need to be replaced with an appropriate equivalent condition, which might be challenging, and which is mainly the reason why we instead approximate the pressure-gradient force to become negligible once cooling becomes dominant.
Future studies that do deal with this complication might be able to incorporate the role of cooling in this derivation and possibly recover the pressure-driven snowplow.

\section{Notable limits}\label{app:notable_limits}

\begin{figure*}
 \centering
 \includegraphics[width=0.8\linewidth, clip=true]{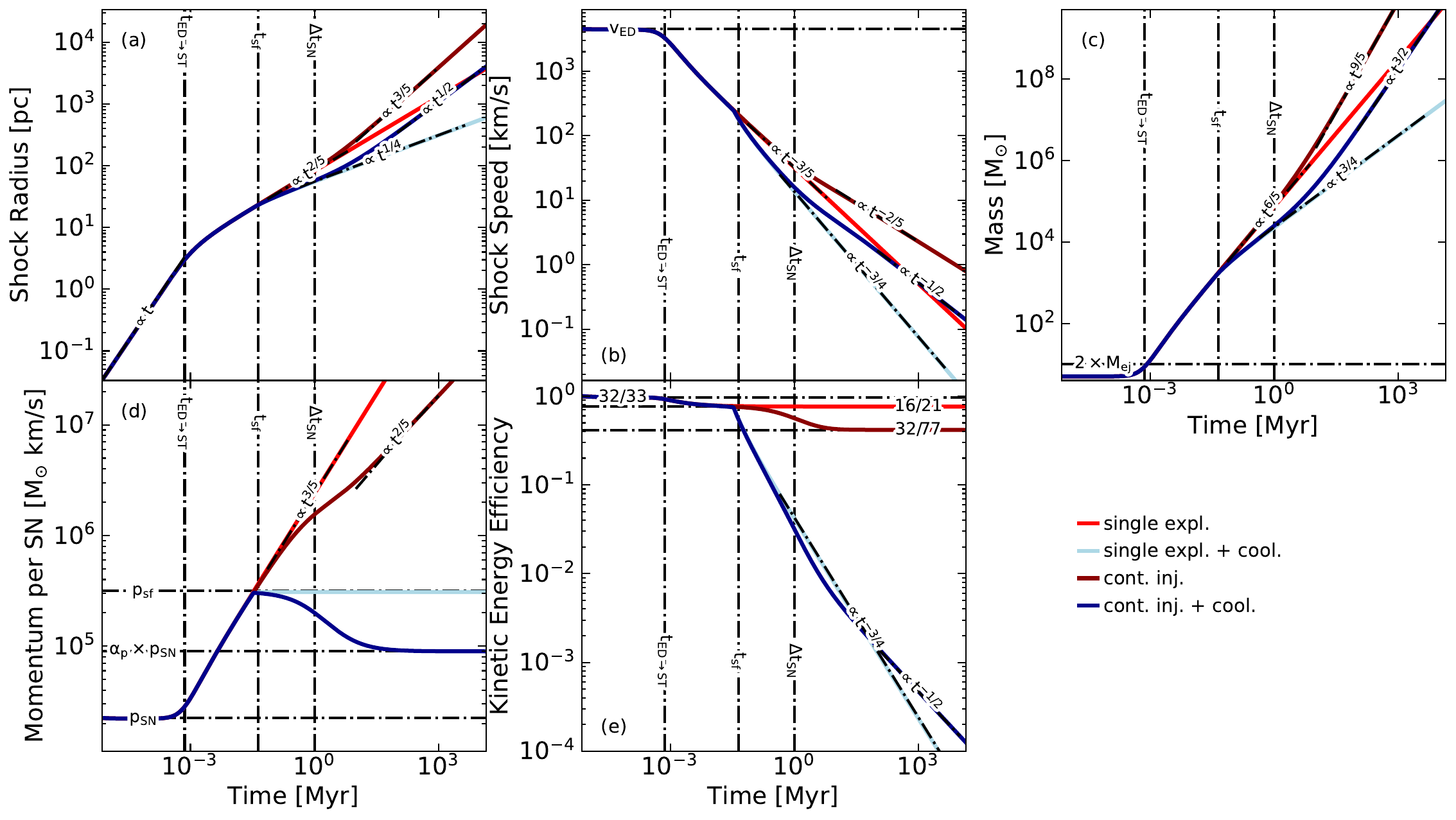}
 \caption{Same as Fig. \ref{fig:stratification} for different models in a stationary, uniform ambient medium. Black, dot-dashed lines denote characteristic scales or limiting behavior described in Appendices~\ref{sec:ED}-\ref{sec:radiative_limit}.} 
 \label{fig:notable_limits}
\end{figure*}

We confirm that our model reproduces the behavior predicted by many previous studies in the appropriate limits.
In Fig.~\ref{fig:notable_limits}, we compare these limits to the results of numerically integrating Eq. \ref{eq:EoM} for  spherically-symmetric blastwave expanding into a stationary, uniform ambient medium.
We consider four different cases:
\begin{enumerate}
    \item A single SN at $t=0$, without radiative cooling.
    \item A single SN at $t=0$, with radiative cooling.
    \item A single SN at $t=0$, followed by continuous injection of energy and mass at a fixed rate, without radiative cooling.
    \item A single SN at $t=0$, followed by continuous injection of energy and mass at a fixed rate, with radiative cooling ($\alpha_{\text{p}} = 4$).
\end{enumerate}

Figure~\ref{fig:notable_limits} shows the temporal evolution of the shock-radius, shock-velocity, swept-up mass, momentum per SN, and the kinetic energy efficiency $f_{\text{kin}} = E_{\text{kin}} / E_{\text{inj}}\left(t\right)$.
We recover all the limiting behavior highlighted below.
Also highlighted are various characteristic timescales, such as the timescale for the transition from the ejecta-dominated to the adiabatic phase $t_{\text{ED}\rightarrow\text{ST}}$\ \citep{1999ApJS..120..299T} and the shell-formation timescale $t_{\text{sf}}$\ \citep{1988ApJ...334..252C, 2015ApJ...802...99K}, which match the times at which the dynamics transition in our model quite well.
The relatively high kinetic energy efficiencies ($> 0.22-0.27$ \citep{1959sdmm.book.....S, 1977ApJ...218..377W}) in the adiabatic cases are a well-known shortcoming of the thin-shell approximation \citep[e.g.][]{1990ApJ...354..513K}.

We note that in many cases, by changing the variable of integration from time to the radius, Eqs. \ref{eq:mass_cons.} - \ref{eq:shell_position} admit analytical solutions in terms of special functions.
The resulting solutions are so-called unified solutions \citep{1999ApJS..120..299T}, which naturally interpolate between the limiting cases.

\subsection{Ejecta-dominated phase}\label{sec:ED}

Shortly after the central point-explosion, the SNR expands with an almost constant speed until the swept-up mass becomes comparable to the mass of the ejecta \citep{1999ApJS..120..299T}.
In our model this behavior is recovered since for small radii $r_{\text{s}} \ll \left(M / \rho_{0}\right)^{1/3}$ the acceleration is dominated by terms in $\Delta P\,r_{\text{s}}^2$ that are $\propto r_{\text{s}}^{-1}$, i.e.
\begin{equation}
    \dot{v}_{\text{s}} \sim \frac{c_{0}}{r_{\text{s}}} \left[c_{1} \frac{E}{M} + \left(k_{\rho} - c_{2}\right) v_{\text{s}}^2\right] ~,
\end{equation}
which rapidly drives $v_{\text{s}}$ towards the constant speed $v_{\text{ED}} = \sqrt{c E / M}$, where $c=c_{1}/\left(c_{2} - k_{\rho}\right)$.

\subsection{Adiabatic phase}\label{sec:ST}

Once $r_{\text{s}} \gtrsim \left(M / \rho_{0}\right)^{1/3}$ the contribution of the ram-pressure-terms in Eq. \ref{eq:EoM} can no longer be neglected.
After a short transitionary period the solution approaches a powerlaw solution resembling the Sedov-Taylor-solution $r_{\text{s}} = \xi_{\text{ST}} \left(E t^2/\rho_{0}\right)^{1/5}$~\citep{1959sdmm.book.....S} for a single explosion, or an energy-driven wind solution $r_{\text{s}} = \xi_{\text{W}} \left(\dot{E} t^3/\rho_{0}\right)^{1/5}$ \citep{1977ApJ...218..377W} for continuous energy injection.
We note that for a better comparison with the spherically symmetrical models of~\citet{1959sdmm.book.....S} and~\citet{1977ApJ...218..377W}
in these expressions $E$ refers to the total energy deposited over the whole sky, i.e. $E = \int \left(\text{d}E/\text{d}\Omega\right) \text{d}\Omega$.

By plugging the respective expressions for $r_{\text{s}}\left(t\right)$ into Eq. \ref{eq:EoM} and solving for $\xi$ we can find the asymptotic solution in the respective regime.
For the Sedov blastwave ($E = const.$) we find
\begin{equation}
    \xi_{\text{ST}} = \left\{\frac{25 c_{0} c_{1}}{8\pi\left[1 - 2 c_{0} \left(3 - (c_{2} - k_{\rho})/3\right)\right]}\right\}^{1/5} ~,
\end{equation}
which for $\gamma=5/3$ and $k_{\rho} = 0$ differs from the analytical solution only by $\lesssim 2.34\,\%$.
For the continuously driven wind we obtain
\begin{equation}
    \xi_{\text{W}} = \left\{\frac{25 c_{0} c_{1}}{4\pi\left[7 - 9 c_{0} \left(3 - (c_{2} - k_{\rho})/3\right)\right]}\right\}^{1/5} ~,
\end{equation}
which matches the solution found by \citet{1977ApJ...218..377W} for $\gamma=5/3$ and $k_{\rho} = 0$ with $1\,\%$ accuracy.

\subsection{Radiative phase}\label{sec:radiative_limit}

In the absence of the pressure-gradient force and any external forces, the combined momentum of the blastwave and the ejecta is conserved as can be seen by integrating the EoM once:
\begin{equation}
   \vec{p} =  M \vec{v}_{s} = \vec{p}_{\text{sf}} + \dot{\vec{p}}_{\text{in}} t ~,
\end{equation}
where $p_{\text{sf}}$ is the momentum at the beginning of the radiative stage, also known as shell formation~\citep[see e.g.][]{2015ApJ...802...99K, 2022ApJS..262....9O, 2024ApJ...965..168R}. 

For a single explosion ($\dot{\vec{p}}_{\text{in}} = 0$) the solution approaches the well-known momentum-conserving snowplow solution~\citep{1988ApJ...334..252C}
\begin{equation}
    r_{\text{s}} \rightarrow \left(\frac{3 p_{\text{sf}} \,t}{\pi\rho_{0}}\right)^{1/4} ~,
\end{equation}
while for continuously driven superbubbles ($\dot{\vec{p}}_{\text{in}} > 0$) the solution approaches that of a momentum-driven wind \citep{2022ApJS..262....9O, 2024ApJ...970...18L}
\begin{equation}
    r_{\text{s}} \rightarrow \left(\frac{3 \dot{p}_{\text{in}} \,t^2}{2\pi\rho_{0}}\right)^{1/4} ~.
\end{equation}
As for the adiabatic case, these results are stated in terms of the total momentum (injection-rate) over the whole sky.

Radiative SNRs are said to merge with the ISM once they have slowed down to the velocity dispersion of the ISM. The merging phase hosts rich phenomenology, such as SNR implosion \citep{1992ApJ...392..131S, 2024ApJ...965..168R}, which requires a combined treatment of the bubble and the shell and is therefore outside of the model presented here.

The thin-shell approximation is a one-zone representation of SNR evolution and therefore by definition is unsuited for describing the interplay of the cold shell and the hot bubble.
In order to describe e.g. the emergence of the reverse shock, the pressure-driven snowplow phase, or SN implosion after merging, a multi-zone extension to this model would be required, which explicitly models the coupled dynamics of the bubble-shell system.
Such multi-zone calculations are outside the scope of this current work.

\section{Role of the cooling threshold}\label{app:cooling_threshold}

\begin{figure*}
 \centering
 \includegraphics[width=0.8\linewidth, clip=true]{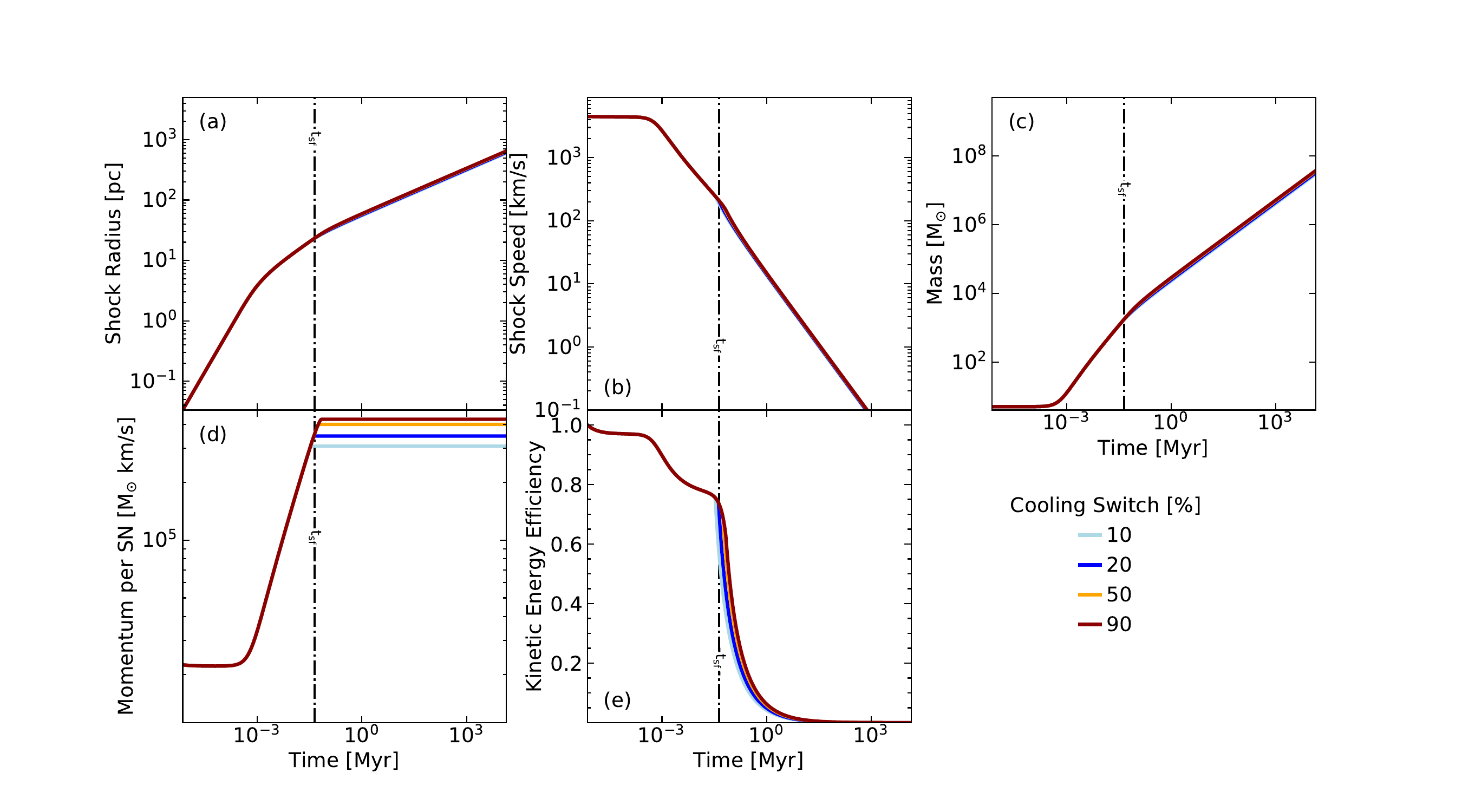}
 \caption{Same as Fig. \ref{fig:stratification} for a single SN in a stationary, uniform ambient medium with different values of the cooling threshold.} 
 \label{fig:cooling_threshold}
\end{figure*}

We appreciate how sensitive our results are to the choice of the cooling threshold, which determines the exact time when the blastwaves transition from the adiabatic to the radiative phase. 
In Fig.~\ref{fig:cooling_threshold}, we show the results of the calculation of a single SNe exploding in a uniform ambient medium for four different values of the threshold ranging from 10 \% (fiducial value) to 90 \%. It is worth noting that for values ($\gtrsim 25 \%$) the sum of the energy that is carried away by radiation up to that point and the kinetic energy (which is approximately constant $\sim 0.75 \, E_{0}$) during the transition) exceeds the total injected energy. 

We find that the dynamics remain largely unchanged with a slight increase of up to $\sim 25 \%$ in the terminal momentum in the cases with a high cooling threshold and a correspondingly slightly higher kinetic energy. 
Since the dynamics are otherwise only marginally affected, we find that the choice of this parameter is largely inconsequential.

\section{Dynamical evolution of tangent vectors}\label{app:tangent_vectors}

We derive an equation of motion for the tangent vectors by switching the order of differentiation
\begin{equation}\label{eq:tangent_evolution}
    \dot{\vec{\partial}}_{i} = \partial_{i} \dot{\vec{r}}_{\text{s}} = \partial_{i} \vec{v}_{\text{s}} + \vec{\nabla} \vec{v}_{\text{ext}} \cdot \vec{\partial}_{i},
\end{equation}
where the external velocity gradient $\vec{\nabla} \vec{v}_{\text{ext}}$ is known a priori, and $\partial_{i} \vec{v}_{\text{s}}$ is evolved along the flow making use of
\begin{equation}
    \frac{\text{d}}{\text{d}t} \partial_{i} \vec{v}_{\text{s}} = \partial_{i} \dot{\vec{v}}_{\text{s}}\left(\vec{r}_{\text{s}}, \vec{v}_{\text{s}}\right),
\end{equation}
evaluated by approximating the local on-surface acceleration gradient using finite differences.
In particular we evaluate the acceleration from eq. \ref{eq:EoM} by shifting the position, velocity and surface are element by the tiny increments
\begin{equation}
    \vec{r}_{s} \rightarrow \vec{r}_{s} \pm \epsilon \vec{\partial}_{i} \qquad \vec{v}_{s} \rightarrow \vec{v}_{s} \pm \epsilon \partial_{i} \vec{v}_{\text{s}} \qquad \vec{\Sigma} \rightarrow \vec{\Sigma} \pm \epsilon \left\lVert\vec{\Sigma}\right\rVert\hat{e}_{i},
\end{equation}
where $\hat{e}_{i} = \vec{\partial}_{i} / \left\lVert\vec{\partial}_{i}\right\rVert$, and we approximated the local curvature radius with the length of the tangential vector for closure. We assume that all other quantities ($M$, $E$, etc.) remain unchanged by a small shift along the surface and we choose $\epsilon \sim 10^{-8}$ for it to be sufficiently small but not so small as to be dominated by numerical noise.

\section{Shape tensor}\label{app:geometry}

We define the shape tensor as
\begin{equation}
    S_{ij} = V_{\text{SNR}}^{-1}\int_{\text{SNR}} \left(\left\lVert \mathbf{x}\right\rVert^2\delta_{ij} - x_{i}x_{j}\right)\text{d}^3\mathbf{x} ~.
\end{equation}
By assuming an approximately ellipsoidal shape, the three ellipsoidal radii, are defined by
\begin{equation}
    r_{i} = \sqrt{2.5\left(\text{tr}\left(S\right) - 2S_{i}\right)} ~,
\end{equation}
where $S_{i}$ are the eigenvalues of $S_{ij}$ and $\text{tr}\left(S\right)$ is the trace.
The smallest, intermediate and largest eigenvalues correspond the minor $a$, semi-major $b$ and major $c$ axis, respectively.
The effective size of an SNR is the geometric mean of the three eigenvalues
\begin{equation}
    r_{\text{eff}} = \left(a b c\right)^{1/3}~.
\end{equation}

To determine the alignment of the an SNR with respect to the galaxy, we measured the pitch angle $\alpha$ of the major axis. As in the case of the shearing sphere, the semi-major axis aligns with the vertical direction and the minor axis is simply shifted by $90^{\circ}$ relative to the major axis.
The pitch angle is defined relative to the direction of galactic rotation, with $\alpha = 90^\circ$ and $\alpha = -90^\circ$ corresponding to the galactic center and anti-center, respectively.
\end{appendix}
\end{document}